\documentclass[prd,twocolumn,floatfix,amsmath,nofootinbib,amssymb,floatfix]{revtex4}
\usepackage{graphicx,color,dcolumn,booktabs,bm}
\usepackage{longtable,lscape}
\usepackage{pdfpages}
\usepackage{txfonts}
\usepackage{overpic}
\usepackage{amssymb}
\usepackage{makecell}
\usepackage{indentfirst}
\usepackage{feynmf}   
\usepackage{slashed}  
\usepackage{cases}
\usepackage{color}
\usepackage{multirow}
\usepackage{threeparttable}
\usepackage{epstopdf}
\usepackage{enumerate}
\usepackage{subfigure}
\usepackage{diagbox}
\usepackage{graphicx,color,dcolumn,booktabs,bm}
\usepackage{mathrsfs}
\usepackage{cancel}
\usepackage{float}
\usepackage[misc]{ifsym}
\usepackage[colorlinks,
            citecolor=blue,
            anchorcolor=red,
            menucolor=red,
            linkcolor=red,
            filecolor=red,
            runcolor=red,
            urlcolor=blue,
            frenchlinks=red]{hyperref}

\begin{document}
\title{Analysis of the strong vertices of $\Sigma_{c}\Delta D^{*}$ and $\Sigma_{b}\Delta B^{*}$ in QCD sum rules}
\author{Jie Lu$^{1,2}$}
\author{Guo-Liang Yu$^{1,2}$}
\email{yuguoliang2011@163.com}
\author{Zhi-Gang Wang$^{1}$}
\email{zgwang@aliyun.com}
\author{Bin Wu$^{1}$}

\affiliation{$^1$ Department of Mathematics and Physics, North China
Electric Power University, Baoding 071003, People's Republic of
China\\$^2$ Hebei Key Laboratory of Physics and Energy Technology, North China Electric Power University, Baoding 071000, China}
\date{\today }

\begin{abstract}
 In this work, we analyze the strong vertices $\Sigma_{c}\Delta D^{*}$ and $\Sigma_{b}\Delta B^{*}$ using the three-point QCD sum rules under the tensor structures $i\epsilon^{\rho\tau\alpha\beta}p_{\alpha}p'_{\beta}$, $p^{\rho}p'^{\tau}$ and $p^{\rho}p^{\tau}$. We firstly calculate the momentum dependent strong coupling constants $g(Q^{2})$ by considering contributions of the perturbative part and the condensate terms $\langle\overline{q}q\rangle$, $\langle g_{s}^{2}GG \rangle$, $\langle\overline{q}g_{s}\sigma  Gq\rangle$ and $\langle\overline{q}q\rangle^{2}$. By fitting these coupling constants into analytical functions and extrapolating them into time-like regions, we then obtain the on-shell values of strong coupling constants for these vertices. The results are $g_{1\Sigma_{c}\Delta D^{*}}=5.13^{+0.39}_{-0.49}$ GeV$^{-1}$, $g_{2\Sigma_{c}\Delta D^{*}}=-3.03^{+0.27}_{-0.35}$ GeV$^{-2}$, $g_{3\Sigma_{c}\Delta D^{*}}=17.64^{+1.51}_{-1.95}$ GeV$^{-2}$, $g_{1\Sigma_{b}\Delta B^{*}}=20.97^{+2.15}_{-2.39}$ GeV$^{-1}$, $g_{2\Sigma_{b}\Delta B^{*}}=-11.42^{+1.17}_{-1.28}$ GeV$^{-2}$ and $g_{3\Sigma_{b}\Delta B^{*}}=24.87^{+2.57}_{-2.82}$ GeV$^{-2}$. These strong coupling constants are important parameters which can help us to understand the strong decay behaviors of hadrons.
\end{abstract}


\maketitle

\section{Introduction}\label{sec1}

The physics of charmed hadrons became an interesting subjects since the observations of $J/\psi$ meson \cite{SLAC-SP-017:1974ind,E598:1974sol}and charmed baryons ($\Lambda_{c},\Sigma_{c}$) \cite{Cazzoli:1975et}. Up to now, lots of charmed baryons have been discovered by different experimental collaborations\cite{ParticleDataGroup:2022pth}. Moreover, many bottom baryons such as $\Lambda_{b}$, $\Xi_{b}$, $\Sigma_{b}$, $\Sigma_{b}^{*}$ and $\Omega_{b}$ have also been confirmed in experiments by CFD and LHCb collaborations\cite{Basile:1981wr,CDF:2007oeq,LHCb:2012kxf,LHCb:2014nae,LHCb:2016rja,LHCb:2018vuc}. Although scientists have devoted much of their energy to this field, but the details of some charmed and bottom baryons are still less known. Thus, many experimental plans for the research of charmed and bottom baryons have been proposed by $\mathrm{\bar{P}}$ANDA\cite{Wiedner:2011mf}, J-PARC\cite{Shirotori:2014nua} and many other facilities. Under this circumstance, theoretical research on production of the baryons is very interesting and important. The strong coupling constants of baryons is an important input parameter which can help us to understand their production and decay processes\cite{Khodjamirian:2011sp}. This is the first motivation for us to carry out the present work.

\begin{figure}[htbp]
\centering
\subfigure[]{\includegraphics[width=4cm]{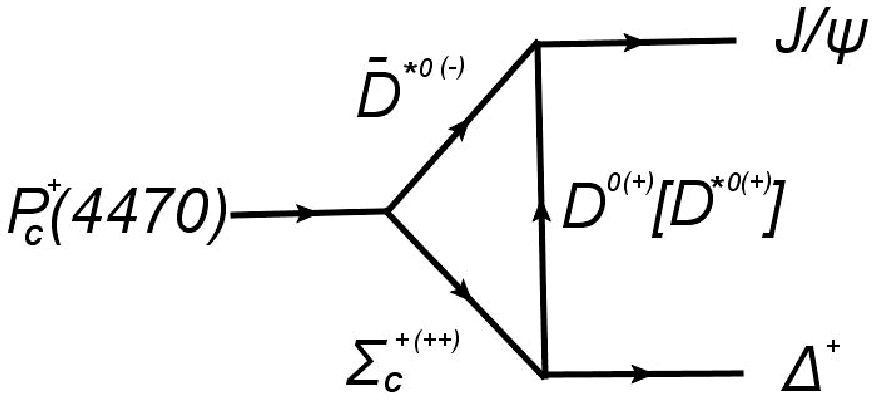}}
\subfigure[]{\includegraphics[width=4cm]{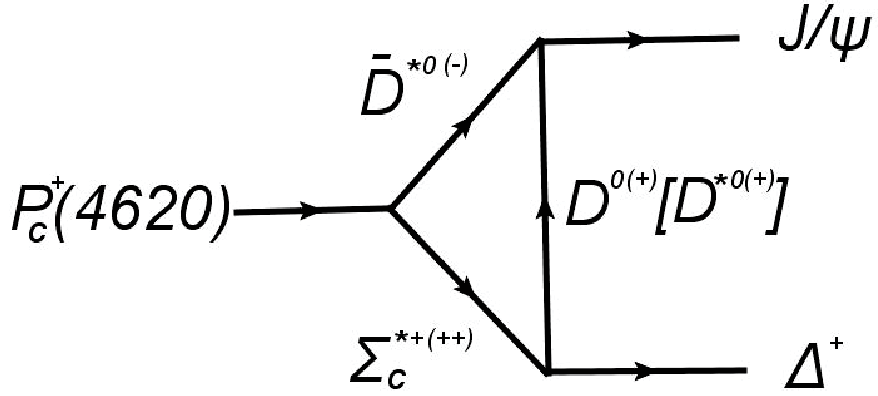}}
\caption{Feynman diagrams for decays: (a)$P_{c}(4470)\to J/\psi\Delta$, (b)$P_{c}(4620)\to J/\psi\Delta$.}
\label{TD}
\end{figure}

Since the observation of X(3872) by Belle collaboration in 2003\cite{Belle:2003nnu}, exotic hadrons which are beyond the usual quark-model emerged like bamboo shoots after a spring rain \cite{BaBar:2003oey,BaBar:2005hhc,Belle:2004lle,BaBar:2006gsq,CDF:2009jgo,Belle:2009rkh,Belle:2011aa,Xiao:2013iha,LHCb:2014zfx,Belle:2014nuw,LHCb:2015yax,CDF:2011pep,LHCb:2017iph,LHCb:2019kea}. Some exotic states were interpreted as hadronic molecular states because their masses are close to the known two-hadrons thresholds\cite{Guo:2017jvc}. However, the study of mass spectra is insufficient to understand the inner structure of these exotic states. We need to further study their strong decay behaviours, where the strong coupling constants are particularly important. For examples, in Ref\cite{Wang:2022ltr}, the authors predicted two pentaquark molecular states $\bar{D}^{*}\Sigma_{c}$ and $\bar{D}^{*}\Sigma^{*}_{c}$ with the QCD sum rules. These two states were named as $P_{c}(4470)$ and $P_{c}(4620)$ which have the isospin $I=\frac{3}{2}$. If we studied their two-body strong decay $P_{c}(4470/4620)\to J/\psi\Delta$, this process can be described by the triangle diagram in Fig. \ref{TD}. From this figure, we can see that analysis of strong vertices $P_{c}\Sigma_{c}D^{*}$, $P_{c}\Sigma_{c}^{*}D^{*}$, $DD^{*}J/\psi$, $D^{*}D^{*}J/\psi$, $\Sigma_{c}\Delta D$, $\Sigma_{c}^{*}\Delta D$, $\Sigma_{c}\Delta D^{*}$ and $\Sigma_{c}^{*}\Delta D^{*}$ is essential for us to study the strong decay behaviors of these two exotic states. This constituents the second motivation of our present work.

The strong interaction between the hadrons is non-perturbative in the low energy region, which can not be studied from the QCD first principle. But, as an important parameter, the strong coupling constant is urgently needed in studying the production and strong decay process of hadrons. Thus, some phenomenological methods are employed to analyze the strong vertices\cite{Navarra:1998vi,Navarra:1999pq,Bracco:1999xe,Khodjamirian:2011jp,Azizi:2014bua,Azizi:2015tya,Azizi:2015jya,Yu:2016pyo,Aliev:2016bhd,Yu:2017ndf,Yu:2018hnv,Aliev:2020aon,Olamaei:2020bvw,Rostami:2020euc,Aliev:2021hqq}.
The QCD sum rules (QCDSR)\cite{Shifman:1978by} and the light-cone sum rules (LCSR) are powerful phenomenological methods to study the strong interaction. In recent years, some coupling constants have been analyzed with LCSR by considering the higher-order QCD corrections and subleading power contributions\cite{Li:2020rcg,Khodjamirian:2020mlb}. These studies show that considering the higher-order QCD corrections and subleading power contributions is very important for the accuracy of the results. In our previous work, we have analyzed the strong vertices $\Sigma_{c}ND$, $\Sigma_{b}NB$, $\Sigma_{c}^{*}ND$, $\Sigma_{b}^{*}NB$, $\Sigma_{c}ND^{*}$ and $\Sigma_{b}NB^{*}$ in the frame work of QCDSR basing on three-point correlation function\cite{Yu:2016pyo,Yu:2017ndf,Yu:2018hnv}, where the higher-order perturbative corrections were neglected. As a continuation of these works, we analyze the strong vertices $\Sigma_{c}\Delta D^{*}$ and $\Sigma_{b}\Delta B^{*}$ using the three-point QCDSR under the tensor structure $i\epsilon^{\rho\tau\alpha\beta}p_{\alpha}p'_{\beta}$, $p^{\rho}p'^{\tau}$ and $p^{\rho}p^{\tau}$. According to our previous work, it showed that the subleading power contributions are really important for the final results. Considering higher-order corrections should make the final results more accurate, however it will also make the calculations of the three-point QCDSR very complicated. Thus, we neglect contributions from these corrections in the present work.

The layout of this paper is as follows. After the introduction in Sec. \ref{sec1}, the strong coupling constants of the vertices $\Sigma_{c}\Delta D^{*}$ and $\Sigma_{b}\Delta B^{*}$ are analyzed by QCD sum rules in Sec. \ref{sec2}. In these analyses, the off-shell cases of the vector mesons are considered. In the QCD side, the perturbative contribution and vacuum condensate terms $\langle\overline{q}q\rangle$, $\langle g_{s}^{2}GG \rangle$, $\langle\overline{q}g_{s}\sigma  Gq\rangle$ and $\langle\overline{q}q\rangle^{2}$ are also considered. In Sec. \ref{sec3}, we present the numerical results and discussions. Sec. \ref{sec4} is reserved for our conclusions. Some calculation details and important formulas are shown in Appendix A and B.

\section{The QCD sum rules for vertices $\Sigma_{c}\Delta D^{*}$ and $\Sigma_{b}\Delta B^{*}$}\label{sec2}

The first step to analyze strong coupling constants with QCD sum rules is to write the following three-point correlation,
\begin{eqnarray}\label{eq1}
\notag
\Pi _{\mu \nu }(p,p',q) &&= {i^2}\int {{d^4}x} \int {{d^4}y} {e^{ - ipx}}{e^{ip'y}} \\
&&\times \langle { 0 | T[{{J^\mu_\Delta }(y){J^\nu_{{D^*}[{B^*}]}}(0){{\bar J }_{{\Sigma _c}[{\Sigma _b}]}}(x)}]| 0} \rangle
\end{eqnarray}
where $T$ is the time ordered product, and $J^{\mu}_{\Delta}$, $J^{\nu}_{D^{*}[B^{*}]}$, $\bar{J}_{\Sigma_{c}[\Sigma_{b}]}$ denote the hadronic interpolating currents of $\Delta$, $D^{*}[B^{*}]$ and $\Sigma_{c}[\Sigma_{b}]$, respectively. These interpolating currents can be expressed as\cite{Ioffe:1981kw},
\begin{eqnarray}\label{eq2}
\notag
&&{J^\mu_\Delta}(y) = {\varepsilon _{ijk}}{{u^{iT}}(y)C{\gamma _\mu }{u^{j}}(y)}{d^{k}}(y)\\
\notag
&&{J^\nu_{{D^*[B^*]}}}(0) = \bar u (0){\gamma _\nu }c[b](0)\\
\notag
&&{J_{\Sigma _c[\Sigma_b] }}(x) = {\varepsilon _{ijk}}{{u^{iT}}(x)C{\gamma _\alpha }{d^j}(x)}{\gamma _5}{\gamma _\alpha }{c[b]^k}(x) \\
&&{\bar {J}_{\Sigma _c[\Sigma_b] }}(x)={J^+_{\Sigma _c[\Sigma_b] }}(x)\gamma_4
\end{eqnarray}
where $i$, $j$ and $k$ represent the color indices and $C$ denotes the charge conjugation operator.

The correlation function can be handled at both hadron and quark level in the framework of QCD sum rules, where the former is called the phenomenological side and the later is called the QCD side. Matching the calculation of these two sides by quark hadron duality, the sum rules for the strong coupling constants can be obtained.

\subsection{The phenomenological side}\label{sec2.1}

In the phenomenological side, a complete sets of hadron states with the same quantum numbers as the hadronic interpolating currents are inserted into the correlation function. After isolating the contributions of ground and excited states, the expression of the correlation function can be written as\cite{Bracco:2011pg},
\begin{widetext}
\begin{eqnarray}\label{eq3}
\notag
\tilde \Pi _{\rho \tau}^{phy}(p,p',q) &&= \Big( {{g^{\mu \rho }} - \frac{{p{'^\mu }p{'^\rho }}}{{p{'^2}}}} \Big)\Big( {{g^{\nu \tau }} - \frac{{{q^\nu }{q^\tau }}}{{{q^2}}}} \Big)\Pi _{\mu \nu }^{phy}(p,p'q) = \Big( {{g^{\mu \rho }} - \frac{{p{'^\mu }p{'^\rho }}}{{p{'^2}}}} \Big)\Big( {{g^{\nu \tau }} - \frac{{{q^\nu }{q^\tau }}}{{{q^2}}}} \Big)\\
\notag
&& \times \Big[\frac{{\left\langle {\left. 0 \right|J_\Delta ^\mu (0)\left| {\Delta \left( {p',s'} \right)} \right.} \right\rangle \left\langle {\left. 0 \right|J_{{D^*}[{B^*}]}^\nu (0)\left| {{D^*}[{B^*}]\left( q \right)} \right.} \right\rangle \left\langle {\Delta \left( {p',s'} \right){D^*}[{B^*}]\left( q \right)\left| {{\Sigma _c}[{\Sigma _b}]\left( {p,s} \right)} \right.} \right\rangle \left\langle {\left. {{\Sigma _c}[{\Sigma _b}]\left( {p,s} \right)} \right|{{\bar J}_{{\Sigma _c}[{\Sigma _b}]}}(0)\left| 0 \right.} \right\rangle }}{{\left( {{p^2} - m_{{\Sigma _c}[{\Sigma _b}]}^2} \right)\left( {{q^2} - m_{{D^*}[{B^*}]}^2} \right)\left( {p{'^2} - m_\Delta ^2} \right)}}\\
\notag
&& + \frac{{\left\langle {\left. 0 \right|J_\Delta ^\mu (0)\left| {N\left( {p',s'} \right)} \right.} \right\rangle \left\langle {\left. 0 \right|J_{{D^*}[{B^*}]}^\nu (0)\left| {{D^*}[{B^*}]\left( q \right)} \right.} \right\rangle \left\langle {N\left( {p',s'} \right){D^*}[{B^*}]\left( q \right)\left| {{\Sigma _c}[{\Sigma _b}]\left( {p,s} \right)} \right.} \right\rangle \left\langle {\left. {{\Sigma _c}[{\Sigma _b}]\left( {p,s} \right)} \right|{{\bar J}_{{\Sigma _c}[{\Sigma _b}]}}(0)\left| 0 \right.} \right\rangle }}{{\left( {{p^2} - m_{{\Sigma _c}[{\Sigma _b}]}^2} \right)\left( {{q^2} - m_{{D^*}[{B^*}]}^2} \right)\left( {p{'^2} - m_N^2} \right)}}\\
\notag
&& + \frac{{\left\langle {\left. 0 \right|J_\Delta ^\mu (0)\left| {\Delta \left( {p',s'} \right)} \right.} \right\rangle \left\langle {\left. 0 \right|J_{{D^*}[{B^*}]}^\nu (0)\left| {D[B]\left( q \right)} \right.} \right\rangle \left\langle {\Delta \left( {p',s'} \right)D[B]\left( q \right)\left| {{\Sigma _c}[{\Sigma _b}]\left( {p,s} \right)} \right.} \right\rangle \left\langle {\left. {{\Sigma _c}[{\Sigma _b}]\left( {p,s} \right)} \right|{{\bar J}_{{\Sigma _c}[{\Sigma _b}]}}(0)\left| 0 \right.} \right\rangle }}{{\left( {{p^2} - m_{{\Sigma _c}[{\Sigma _b}]}^2} \right)\left( {{q^2} - m_{D[B]}^2} \right)\left( {p{'^2} - m_\Delta ^2} \right)}}\\
\notag
&& + \frac{{\left\langle {\left. 0 \right|J_\Delta ^\mu (0)\left| {N\left( {p',s'} \right)} \right.} \right\rangle \left\langle {\left. 0 \right|J_{{D^*}[{B^*}]}^\nu (0)\left| {D[B]\left( q \right)} \right.} \right\rangle \left\langle {N\left( {p',s'} \right)D[B]\left( q \right)\left| {{\Sigma _c}[{\Sigma _b}]\left( {p,s} \right)} \right.} \right\rangle \left\langle {\left. {{\Sigma _c}[{\Sigma _b}]\left( {p,s} \right)} \right|{{\bar J}_{{\Sigma _c}[{\Sigma _b}]}}(0)\left| 0 \right.} \right\rangle }}{{\left( {{p^2} - m_{{\Sigma _c}[{\Sigma _b}]}^2} \right)\left( {{q^2} - m_{D[B]}^2} \right)\left( {p{'^2} - m_N^2} \right)}}\\
&& + h.c.\Big ]
\end{eqnarray}
where $h.c.$ denotes the contributions of higher resonances and continuum states. From this above equation, we can see that the current $J_{\Delta}^{\mu}(0)$ couples not only with the baryon $J^{P}=\frac{3}{2}^{+}$ but also with the state of $\frac{1}{2}^{+}$. Similarly, the meson current $J^{\nu}_{D^{*}[B^{*}]}(0)$ couples with both the vector meson with $J^{P}=1^{-}$ and the pseudoscalar meson with $J^{P}=0^{-}$. Therefore, there will be some redundant terms, that is the second, third and fourth term in Eq. (\ref{eq3})). They will disturb the items that we are interested in(the first term in Eq. (\ref{eq3})). These redundant matrix elements can be parameterized by the following equations,

\begin{eqnarray}\label{eq4}
\notag
\langle {\left. 0 \right|J_\Delta ^\mu (0)\left| {N(p',s')} \right.}\rangle  =&& {\lambda _N}{U_N}(p',s')p{'_\mu }\\
\notag
\langle {\left. 0 \right|J_{D*[B*]}^\nu (0)\left| {D[B]\left( q \right)} \right.} \rangle  =&& {m_{D[B]}}{f_{D[B]}}{q_\mu }\\
\notag
\left\langle {\Delta \left( {p',s'} \right)D[B]\left( q \right)\left| {{\Sigma _c}[{\Sigma _b}]\left( {p,s} \right)} \right.} \right\rangle =&& g{{\bar U}_\alpha }({p',s'}){q_\alpha }{U_{{\Sigma _c}[{\Sigma _b}]}}({p,s})\\
\notag
\left\langle {N\left( {p',s'} \right)D[B]\left( q \right)\left| {{\Sigma _c}[{\Sigma _b}]({p,s})} \right.} \right\rangle =&&g'{{\bar U}_N}({p',s'})i{\gamma _5}{U_{{\Sigma _c}[{\Sigma _b}]}}({p,s}) \\
\left\langle {N\left( {p',s'} \right){D^*}[{B^*}]\left( q \right)\left| {{\Sigma _c}[{\Sigma _b}]\left( {p,s} \right)} \right.} \right\rangle =&& {{\bar U}_N}({p',s'})[ f_1^{}{\gamma _\beta }-\frac{{f_2^{}{\sigma _{\alpha \beta }}}}{{{m_{{\Sigma _c}[{\Sigma _b}]}} + {m_N}}}{q^\alpha }]{\gamma _5}{U_{{\Sigma _c}[{\Sigma _b}]}}({p,s}){\varepsilon _\beta }
\end{eqnarray}
where $N$ represents baryon with spin parity $\frac{1}{2}^{+}$, $D[B]$ is the pseudoscalar charmed(bottom) meson, $U(p,s)$ and $U_{\alpha}(p,s)$ are the spinor wave functions of the baryon with spin parity $\frac{1}{2}^{+}$ and $\frac{3}{2}^{+}$, respectively, $\varepsilon_{\beta}$ is the polarization vector of the vector meson $D^{*}[B^{*}]$, $\lambda_{N}$ is the pole residues, $f_{D[B]}$ is the decay constant. To extract the contributions of $\Sigma_c$[$\Sigma_b$], $D^{*}[B^{*}]$ and $\Delta$, and eliminate the contaminations of the redundant terms(see Eq. (\ref{eq4})), the projection operators $(g^{\mu\rho}-\frac{p'^{\mu}p'^{\rho}}{p'^{2}})$ and $(g^{\nu\tau}-\frac{q^{\nu}q^{\tau}}{q^{2}})$ are employed in Eq. (\ref{eq3}). The matrix elements about the vertex $\Sigma_{c}[\Sigma_{b}]\Delta D^{*}[B^{*}]$ can be written as follows,

\begin{eqnarray}\label{eq5}
\notag
\left\langle {\left. 0 \right|J_\Delta ^\mu (0) \left| {\Delta\left( {p',s'} \right)} \right.} \right\rangle  &&= \lambda _\Delta ^{}{U_\mu }({p',s'})\\
\notag
\left\langle {\left. {{\Sigma _c}[{\Sigma _b}]\left( {p,s} \right)} \right|{{\bar J}_{{\Sigma _c}[{\Sigma _b}]}} (0)\left| 0 \right.} \right\rangle  &&= \lambda _{{\Sigma _c}[{\Sigma _b}]}\bar U({p,s})\\
\notag
\left\langle {\left. 0 \right|J_{{D^*}[{B^*}]}^\nu (0) \left| {{D^*}[{B^*}]\left( q \right)} \right.} \right\rangle  &&= {m_{{D^*}[{B^*}]}}{f_{{D^*}[{B^*}]}}\varepsilon _\nu ^*\\
\notag
\left\langle {\Delta \left( {p',s'} \right){D^*}[{B^*}]\left( q \right)\left| {{\Sigma _c}[{\Sigma _b}]\left( {p,s} \right)} \right.} \right\rangle  &&= {{\bar U}_\alpha }({p',s'})[g_1( {q_\alpha }{\gamma _\beta } - {g_{\alpha \beta }}\slashed{q})+ g_2( {P_\beta }{q_\alpha }- Pq{g_{\alpha \beta }})\\
&&+ g_3( {q_\alpha }{q_\beta }- {q^2}{g_{\alpha \beta }})]{\gamma _5}{U_{{\Sigma _c}[{\Sigma _b}]}}\left( {p,s} \right){\varepsilon _\beta }
\end{eqnarray}
where $P=p+p'$.

The matrix elements appearing in Eq. (\ref{eq3}) are substituted with Eqs. (\ref{eq4}) and (\ref{eq5}). Then, the correlation function in the phenomenological side can be written as the following form,
\begin{eqnarray}\label{eq6}
\notag
\tilde \Pi _{\rho \tau }^{\mathrm phy}(p,p',q) &&= \frac{1}{{({{p^2} - m_{{\Sigma _c}[{\Sigma _b}]}^2})({{q^2} - m_{{D^*}[{B^*}]}^2})({p{'^2} - m_\Delta ^2})}}({{g^{\mu \rho }} - \frac{{p{'^\mu }p{'^\rho }}}{{p{'^2}}}})({{g^{\nu \tau }} - \frac{{{q^\nu }{q^\tau }}}{{{q^2}}}})\\
\notag
&&\times\lambda _\Delta \lambda _{{\Sigma _c}[{\Sigma _b}]}{m_{{D^*}[{B^*}]}}{f_{{D^*}[{B^*}]}}[{ - ({\slashed p' + {m_\Delta }})({{g_{\mu \alpha }} - \frac{{{\gamma _\mu }{\gamma _\alpha }}}{3} - \frac{{2p{'_\mu }p{'_\alpha }}}{{3m_\Delta ^2}} + \frac{{p{'_\mu }{\gamma _\alpha } - p{'_\alpha }{\gamma _\mu }}}{{3{m_\Delta }}}})}]\\
\notag
&&\times[ {g_1({{q_\alpha }{\gamma _\beta } - {g_{\alpha \beta }}\slashed q}) + g_2({{P_\beta }{q_\alpha } - Pq{g_{\alpha \beta }}}) + g_3({q_\alpha q_\beta - {q^2}{g_{\alpha \beta }}})}]{\gamma _5}({ - {g_{\nu \beta }} + \frac{{{q_\nu }{q_\beta }}}{{{q^2}}}})({\slashed p' + \slashed q + {m_{{\Sigma _c}[{\Sigma _b}]}}}) \\
&& + h.c.
\end{eqnarray}
\end{widetext}

From Eq. (\ref{eq6}), we can see that the correlation function will have so complex tensor structure, \emph{e}.\emph{g}. $\slashed p \slashed p^{\prime}\gamma^\rho \gamma^\tau\gamma^{5}$, $\slashed p\gamma^\rho \gamma^{\tau}\gamma^{5}$, $\slashed p \slashed p^{\prime }g^{\rho \tau}\gamma^{5}$, $p^{\rho}p^{\tau}\gamma^{5}$, $\cdots$ that the calculation become tedious and lengthy.
Theoretically, if all the criteria of QCD sum rules are satisfied, each tensor structure can lead to the same results. For simplicity, we choose the tensor structure in the following ways,
\begin{eqnarray}\label{eq8}
\notag
Tr[\tilde \Pi _{\rho \tau }^{\mathrm phy}(p,p',q)] &&= {{\bar \Pi }_1^{\mathrm phy}}(p,p',q)i{\varepsilon ^{\rho \tau \lambda \delta }}{p^\lambda }p{'^\delta }\\
\notag
Tr[{\gamma _5}\tilde \Pi _{\rho \tau }^{\mathrm phy}(p,p',q)] &&= {{\bar \Pi }_2^{\mathrm phy}}(p,p',q){p^\rho }{p'^\tau } \\
&&+ {{\bar \Pi }_3^{\mathrm phy}}(p,p',q){p^\rho }{p^\tau } + ...
\end{eqnarray}
$\bar{\Pi}^{\mathrm{phy}}_{1}$, $\bar{\Pi}^{\mathrm{phy}}_{2}$ and $\bar{\Pi}^{\mathrm{phy}}_{3}$ are named as scalar invariant amplitudes which can be obtained by using Eq. (\ref{eq6}) and (\ref{eq8}),
\begin{eqnarray}\label{eq9}
\notag
\bar{\Pi}^{\mathrm{phy}}_{1}(p,p',q)&&=\frac{Ag_{1}+Bg_{2}+Cg_{3}}{{({{p^2} - m_{{\Sigma _c}[{\Sigma _b}]}^2})({{q^2} - m_{{D^*}[{B^*}]}^2})({p{'^2} - m_\Delta ^2})}}\\
\notag
\bar{\Pi}^{\mathrm{phy}}_{2}(p,p',q)&&=\frac{Dg_{1}+Eg_{2}+Fg_{3}}{{({{p^2} - m_{{\Sigma _c}[{\Sigma _b}]}^2})({{q^2} - m_{{D^*}[{B^*}]}^2})({p{'^2} - m_\Delta ^2})}} \\
\notag
\bar{\Pi}^{\mathrm{phy}}_{3}(p,p',q)&&=\frac{Gg_{1}+Hg_{2}+Ig_{3}}{{({{p^2} - m_{{\Sigma _c}[{\Sigma _b}]}^2})({{q^2} - m_{{D^*}[{B^*}]}^2})({p{'^2} - m_\Delta ^2})}} \\
\end{eqnarray}
where,
\begin{eqnarray}\label{eq10}
\notag
&&A=\frac{2(m_{\Sigma_{c}[\Sigma_{b}]}^{2}+4m_{\Sigma_{c}[\Sigma_{b}]}m_{\Delta}-m_{D^{*}[B^{*}]}^{2}+3m_{\Delta}^{2})}{3m_{\Delta}}\\
\notag
&&B=-\frac{4(m_{\Sigma_{c}[\Sigma_{b}]}^{2}-m_{\Delta}^{2})}{3}\\
\notag
&&C=-\frac{4m_{D^{*}[B^{*}]}^{2}}{3}\\
\notag
&&D=-\frac{2(m_{\Sigma_{c}[\Sigma_{b}]}^{2}-4m_{\Sigma_{c}[\Sigma_{b}]}m_{\Delta}-m_{D^{*}[B^{*}]}^
{2}+m_{\Delta}^{2})}{3m_{\Delta}} \\
\notag
&&E=-F=-\frac{4(m_{\Sigma_{c}[\Sigma_{b}]}^{2}-2m_{\Sigma_{c}[\Sigma_{b}]}m_{\Delta}-m_{D^{*}[B^{*}]}^{2}+m_{\Delta}^{2})}{3} \\
\notag
&&G=-\frac{4m_{\Delta}}{3} \\
&&H=I=-\frac{4(m_{\Sigma_{c}[\Sigma_{b}]}^{2}-2m_{\Sigma_{c}[\Sigma_{b}]}m_{\Delta}-m_{D^{*}[B^{*}]}^{2}+m_{\Delta}^{2})}{3}
\end{eqnarray}

\subsection{The QCD side}\label{sec2.2}

In the QCD side, we firstly contract all of the quark fields in the correlation function with Wick's theorem,
\begin{eqnarray}\label{eq11}
\notag
\Pi _{\mu \nu }^{QCD}(p,p',q) &&=  - 2{i^2}{\varepsilon _{ijk}}{\varepsilon _{i'j'k'}}\int {{d^4}x} {d^4}y{e^{ip'x}}{e^{iqy}}\\
\notag
&&\times [S_d^{k'j}(x){\gamma _\alpha }CS_u^{ii'T}(x)C{\gamma _\mu }S_u^{j'm}(x - y)\\
&&{\gamma _\nu }S_Q^{mk}(y){\gamma ^\alpha }{\gamma _5}]
\end{eqnarray}
Here, $S_{u[d]}^{mn}(x)$ and $S_{c[b]}^{mn}(x)$ are light and heavy quark full propagators which can be written as\cite{Pascual:1984zb,Reinders:1984sr},

\begin{eqnarray}
\notag
{S^{mn}_{u[d]}}(x)&& = \frac{{i\slashed{x}}}{{2{\pi ^2}{x^4}}}{\delta ^{mn}} - \frac{{{m_{u[d]}}}}{{4{\pi ^2}{x^4}}}{\delta ^{mn}} - \frac{{\left\langle {\bar qq} \right\rangle }}{{12}}{\delta ^{mn}} \\
\notag
&&+ \frac{{i\slashed{x}{m_{u[d]}}\left\langle {\bar qq} \right\rangle }}{{48}}{\delta ^{mn}} - \frac{{{x^2}\left\langle {\bar q{g_s}\sigma Gq} \right\rangle }}{{192}}{\delta ^{mn}} \\
\notag
&&+ \frac{{i{x^2}\slashed{x}{m_{u[d]}}\left\langle {\bar q{g_s}\sigma Gq} \right\rangle }}{{1152}}{\delta ^{mn}}+...
\end{eqnarray}
\begin{eqnarray}\label{eq12}
\notag
{S^{mn}_{c[b]}}(x) &&= \frac{i}{{{{(2\pi )}^4}}}\int {{d^4}k} {e^{ - ik \cdot x}}\Big\{ \frac{{{\delta ^{mn}}}}{{\slashed k - {m_{c[b]}}}} \\
\notag
&&- \frac{{{g_s}G_{\alpha \beta }^at_{mn}^a}}{4}\frac{{{\sigma ^{\alpha \beta }}(\slashed{k} + {m_{c[b]}}) + (\slashed{k} + {m_{c[b]}}){\sigma ^{\alpha \beta }}}}{{{{({k^2} - m_{c[b]}^2)}^2}}}  \\
\notag
&&- \frac{{g_s^2{{({t^a}{t^b})}_{mn}}G_{\alpha \beta }^aG_{\mu \nu }^b({f^{\alpha \beta \mu \nu }} + {f^{\alpha \mu \beta \nu }} + {f^{\alpha \mu \nu \beta }})}}{{4{{({k^2} - m_{c[b]}^2)}^5}}} \\
&&+ ...\Big\}
\end{eqnarray}
where $\langle g_{s}^{2}G^{2}\rangle=\langle g_{s}^{2}G^{a}_{\alpha\beta}G^{a\alpha\beta}\rangle$, $D_{\alpha}=\partial_{\alpha}-ig_{s}G^{a}_{\alpha}t^{a}$, $t^{a}=\frac{\lambda^{a}}{2}$, $\lambda^{a}(a=1,..,8)$ are the Gell-Mann matrixes, $i$ and $j$ are color indices, $\sigma_{\alpha\beta}=\frac{i}{2}[\gamma_{\alpha},\gamma_{\beta}]$ and $f^{\lambda\alpha\beta}$, $f^{\alpha\beta\mu\nu}$ have the following forms,
\begin{eqnarray}
{f^{\lambda \alpha \beta }} &&= (\slashed k + {m_c}){\gamma ^\lambda }(\slashed k + {m_c}){\gamma ^\alpha }(\slashed k + {m_c}){\gamma ^\beta }(\slashed k + {m_c})
\end{eqnarray}
\begin{eqnarray}
\notag
{f^{\alpha \beta \mu \nu }} &&= (\slashed k + {m_c}){\gamma ^\alpha }(\slashed k + {m_c}){\gamma ^\beta }(\slashed k + {m_c})\\
&&{\gamma ^\mu }(\slashed k + {m_c}){\gamma ^\nu }(\slashed k + {m_c})
\end{eqnarray}

Taking the same way as the phenomenological side, the correlation function in QCD side can also be written as,
\begin{eqnarray}\label{eq15}
\notag
\tilde \Pi _{\rho \tau }^{\mathrm QCD}(p,p',q) = ({{g^{\mu \rho }} - \frac{{p{'^\mu }p{'^\rho }}}{{p{'^2}}}})({{g^{\nu \tau }} - \frac{{{q^\nu }{q^\tau }}}{{{q^2}}}})\Pi _{\mu \nu }^{\mathrm QCD}(p,p',q)
\end{eqnarray}
and
\begin{eqnarray}\label{eq15}
\notag
Tr[\tilde \Pi _{\rho \tau }^{\mathrm QCD}(p,p',q)] &&= {{\bar \Pi }_1^{\mathrm QCD}}(p,p',q)i{\varepsilon ^{\rho \tau \lambda \delta }}{p^\lambda }p{'^\delta }\\
\notag
Tr[{\gamma _5}\tilde \Pi _{\rho \tau }^{\mathrm QCD}(p,p',q)] &&= {{\bar \Pi }_2^{\mathrm QCD}}(p,p',q){p^\rho }{p'^\tau } \\
&&+ {{\bar \Pi }_3^{\mathrm QCD}}(p,p',q){p^\rho }{p^\tau } + ...
\end{eqnarray}

\begin{figure*}[htbp]
\centering
\includegraphics[width=15cm]{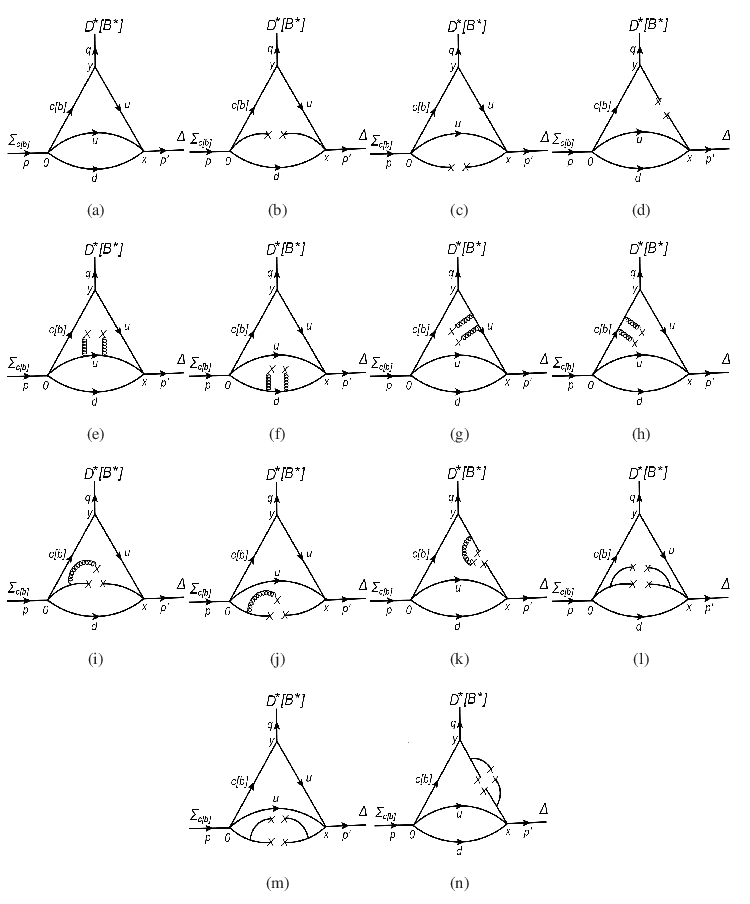}
\caption{Feynman diagrams for the perturbative part and vacuum condensate terms.}
\label{fig:FD}
\end{figure*}

After conducting operator product expansion(OPE) and taking their imaginary part, we can obtain the spectral density of correlation function. Finally, the correlation function can be written as following form by using the dispersion relation,
\begin{eqnarray}
\notag
\bar \Pi _i^{\mathrm{QCD}}(p,p',q) &&=  - \int\limits_{{s_1}}^\infty  {\int\limits_{{u_1}}^\infty  {dsdu} } \\
&& \times \frac{{\bar \rho _i^{\mathrm{pert}}(s,u,{q^2}) + \bar \rho _i^{\mathrm{non - pert}}(s,u,{q^2})}}{{(s - {p^2})(u - p{'^2})}}
\end{eqnarray}
where $s=p^{2}$, $u=p'^{2}$, $q=p-p'$, $s_{1}$ and $u_{1}$ are the kinematic limits which are taken as $(2m_{q}+m_{Q})^{2}$ and $9m_{q}^{2}$ respectively. The QCD spectral density $\bar{\rho}_{i}(s,u,q^{2})$ can be obtained by Cutkosky's rules\cite{Wang:2007ys,MarquesdeCarvalho:1999bqs,Shi:2019hbf,Zhao:2020mod,Wang:2012hu,Yang:2005bv}, and their calculation details are briefly discussed in Appendix A. Full expressions of the QCD spectral density for different tensor structures are shown in Appendix B. The contributions of perturbative part and the vacuum condensation
terms including $\langle\overline{q}q\rangle$, $\langle g_{s}^{2}GG \rangle$, $\langle\overline{q}g_{s}\sigma  Gq\rangle$ and $\langle\overline{q}q\rangle^{2}$ are all considered, where their Feynman diagrams are shown in Fig. \ref{fig:FD}.

\subsection{The strong coupling constants}\label{sec2.3}
We take the change of variables $p^{2}\to-P^{2}$, $p'^{2}\to-P'^{2}$ and $q^{2}\to-Q^{2}$ and perform double Borel transformation\cite{Ioffe:1982ia,Ioffe:1982qb} to both the phenomenological and QCD sides. The variables $P^{2}$ and $P'^{2}$ are replaced by $T_{1}^{2}$ and $T_{2}^{2}$ which are called the Borel parameters. Then we take $T^{2}=T_{1}^{2}$ and $T_{2}^{2}=kT_{1}^{2}=kT^{2}$, where $k=m_{\Delta}^{2}/m_{\Sigma_{c}[\Sigma_{b}]}^2$. Finally, we can obtain the following equations about the strong coupling constants $g_{i}(i=1,2,3)$ using the quark-hadron duality condition,
\begin{widetext}
\begin{eqnarray}\label{eq:17}
\frac{{{\lambda _\Delta }{\lambda _{{\Sigma _c}[{\Sigma _b}]}}{m_{{D^*}[{B^*}]}}{f_{{D^*}[{B^*}]}}}}{{{Q^2} + m_{{D^*}[{B^*}]}^2}}{\begin{bmatrix}
A&B&C\\
D&E&F\\
G&H&I
\end{bmatrix}} \begin{bmatrix}
{g_{1{\Sigma _c}[{\Sigma _b}]\Delta {D^*}[{B^*}]}}({Q^2})\\
{g_{2{\Sigma _c}[{\Sigma _b}]\Delta {D^*}[{B^*}]}}({Q^2})\\
{g_{3{\Sigma _c}[{\Sigma _b}]\Delta {D^*}[{B^*}]}}({Q^2})
\end{bmatrix}{e^{ - \frac{{m_\Delta ^2}}{{k{T^2}}}}}{e^{ - \frac{{m_{{\Sigma _c}[{\Sigma _b}]}^2}}{{{T^2}}}}} = \int\limits_{9m_u^2}^{{u_0}} {du} \int\limits_{{{(2{m_u} + {m_{c[b]}})}^2}}^{{s_0}} {ds}\begin{bmatrix}
{{\bar \rho }_1}(s,u,{Q^2})\\
{{\bar \rho }_2}(s,u,{Q^2})\\
{{\bar \rho }_3}(s,u,{Q^2})
\end{bmatrix}{e^{ - \frac{u}{{k{T^2}}}}}{e^{ - \frac{s}{{{T^2}}}}}
\end{eqnarray}

The momentum dependent coupling constants can be expressed as,

\begin{eqnarray}\label{eq:18}
\notag
\begin{bmatrix}
{g_{1{\Sigma _c}[{\Sigma _b}]\Delta {D^*}[{B^*}]}}({Q^2})\\
{g_{2{\Sigma _c}[{\Sigma _b}]\Delta {D^*}[{B^*}]}}({Q^2})\\
{g_{3{\Sigma _c}[{\Sigma _b}]\Delta {D^*}[{B^*}]}}({Q^2})
\end{bmatrix} = {e^{\frac{{m_\Delta ^2}}{{k{T^2}}}}}{e^{\frac{{m_{{\Sigma _c}[{\Sigma _b}]}^2}}{{{T^2}}}}}\frac{{{Q^2} + m_{{D^*}[{B^*}]}^2}}{{{\lambda _\Delta }{\lambda _{{\Sigma _c}[{\Sigma _b}]}}{m_{{D^*}[{B^*}]}}{f_{{D^*}[{B^*}]}}(CEG - BFG - CDH + AFH + BDI - AEI)}}\\
 \times \{ \int\limits_{9m_u^2}^{{u_0}} {du} \int\limits_{{{(2{m_u} + {m_{c[b]}})}^2}}^{{s_0}} {ds} {e^{ - \frac{u}{{k{T^2}}}}}{e^{ - \frac{s}{{{T^2}}}}} {\begin{bmatrix}
{ - (FH + EI)}&{CH - BI}&{CE - BF}\\
{FG - DI}&{AI - CG}&{CD - AF}\\
{DH - EG}&{BG - AH}&{ - (BD - AE)}
\end{bmatrix}} \begin{bmatrix}{}
{{\bar \rho }_1}(s,u,{Q^2})\\
{{\bar \rho }_2}(s,u,{Q^2})\\
{{\bar \rho }_3}(s,u,{Q^2})
\end{bmatrix}\}
\end{eqnarray}
\end{widetext}
where $s_{0}$ and $u_{0}$ are the threshold parameters which are introduced to eliminate the $h.c.$ terms in Eq. (\ref{eq6}). They satisfy the relations, $m^{2}_{\Sigma_{c}[\Sigma_{b}]}<s_{0}<m'^{2}_{\Sigma_{c}[\Sigma_{b}]}$ and $m^{2}_{\Delta}<u_{0}<m'^{2}_{\Delta}$, where $m$ and $m'$ are the masses of the ground and first excited states of the baryons.
\section{Numerical results and Discussions}\label{sec3}
This section is devoted to analyzing the numerical results of the coupling constants. The masses of the hadrons and quarks used in the present work are taken as the standard values which are adopted from PDG\cite{ParticleDataGroup:2022pth}. Their values are $m_{\Sigma_{c}}=2.45$ GeV, $m_{\Sigma_{b}}=5.81$ GeV, $m_{\Delta}=1.23$ GeV, $m_{D^{*}}=2.01$ GeV and $m_{B^{*}}=5.33$ GeV, $m_{u(d)}=0.006\pm0.001$ GeV, $m_{c}=1.275\pm0.025$ GeV and $m_{b}=4.18\pm0.03$ GeV. The pole residues and decay constants are adopted to be $\lambda_{\Sigma_{c}}=0.045\pm0.015$ GeV$^{3}$\cite{Azizi:2008ui}, $\lambda_{\Sigma_{b}}=0.062\pm0.018$ GeV$^{3}$\cite{Azizi:2008ui}, $\lambda_{\Delta}=0.03\pm0.002$ GeV$^{3}$\cite{Ioffe:1981kw}, $f_{D^{*}}=0.263\pm0.021$ GeV\cite{Wang:2015mxa} and $f_{B^{*}}=0.213\pm0.018$ GeV\cite{Wang:2015mxa}. As for the vacuum condensates, their values are $\langle\overline{q}q\rangle=-(0.23\pm0.01)^{3}$ GeV$^{3}$\cite{ParticleDataGroup:2022pth}, $\langle\overline{q}g_{s}\sigma Gq\rangle=m_{0}^{2}\langle\overline{q}q\rangle$\cite{ParticleDataGroup:2022pth}, $m_{0}^{2}=0.8\pm0.1$ GeV$^2$\cite{Narison:2010cg,Narison:2011xe,Narison:2011rn}, $\langle g_{s}^{2}G^{2}\rangle=0.88\pm0.15$ GeV$^{4}$\cite{Narison:2010cg,Narison:2011xe,Narison:2011rn}, $\langle f^{3}G^{3}\rangle=(8.8\pm5.5)$ GeV$^{2}\langle g_{s}^{2}G^{2}\rangle$\cite{Narison:2010cg,Narison:2011xe,Narison:2011rn}. The threshold parameters $s_{0}$ and $u_{0}$ in Eq. (\ref{eq:20}) are used to eliminate the contributions of the excited and continuum states. They commonly satisfy $s_{0}=(m_{\Sigma_{c}[\Sigma_{b}]}+\delta_{\Sigma_{c}[\Sigma_{b}]})^{2}$ and $u_{0}=(m_{\Delta}+\delta_{\Delta})^2$, where the parameters $\delta_{\Sigma_{c}[\Sigma_{b}]}$ and $\delta_{\Delta}$ are taken as 0.4 $\sim$ 0.6 GeV\cite{Bracco:2011pg}.

In the framework of QCD sum rules, two conditions should also be satisfied, which are the pole dominance and convergence of OPE. To analyze the pole contribution, we write down,

\begin{eqnarray}\label{eq:19}
\notag
\bar\Pi^{\mathrm{QCD}}_{i\mathrm{pole}}(T^{2})&&=-\int_{(2m_{u}+m_{c[b]})^2}^{s_{0}}\int_{9m_{u}^{2}}^{u_{0}}\bar\rho^{\mathrm{QCD}}_{i}(s,u,Q^2)e^{-\frac{s}{T^{2}}}e^{-\frac{u}{kT^{2}}}dsdu \\
\bar\Pi^{\mathrm{QCD}}_{i\mathrm{cont}}(T^{2})&&=-\int_{s_{0}}^{\infty}\int_{u_{0}}^{\infty}\bar\rho^{\mathrm{QCD}}_{i}(s,u,Q^2)e^{-\frac{s}{T^{2}}}e^{-\frac{u}{kT^{2}}}dsdu
\end{eqnarray}
Then, the pole contribution can be defined as\cite{Bracco:2011pg},
\begin{eqnarray}\label{eq:20}
\mathrm{Pole}_{i}=\frac{\bar\Pi^{\mathrm{QCD}}_{i\mathrm{pole}}(T^{2})}{\bar\Pi^{\mathrm{QCD}}_{i\mathrm{pole}}(T^{2})+\bar\Pi^{\mathrm{QCD}}_{i\mathrm{cont}}(T^{2})}
\end{eqnarray}
The convergence of OPE is quantified via the contributions of the vacuum condensates of dimension $n$, which is defined as,
\begin{eqnarray}\label{eq:21}
D(n) = \frac{{\int\limits_{{{(2{m_u} + {m_{c[b]}})}^2}}^{{s_0}} {\int\limits_{9m_u^2}^{{u_0}} {\bar \rho _{i}^{\mathrm{QCD};n}(s,u,{Q^2}){e^{ - \frac{s}{{{T^2}}}}}{e^{ - \frac{u}{{k{T^2}}}}}} duds} }}{{\int\limits_{{{(2{m_u} + {m_{c[b]}})}^2}}^{{s_0}} {\int\limits_{9m_u^2}^{{u_0}} {\bar \rho _i^{\mathrm{QCD}}(s,u,{Q^2}){e^{ - \frac{s}{{{T^2}}}}}{e^{ - \frac{u}{{k{T^2}}}}}} duds} }}
\end{eqnarray}
where $\bar\rho_{i}^{\mathrm{QCD}}(s,u,Q^{2})$ and $\bar\rho_{i}^{\mathrm{QCD},n}(s,u,Q^{2})$ represent the spectral densities of total and the $n$th dimension vacuum condensates, respectively.

\begin{figure}[htbp]
\centering
\subfigure[]{\includegraphics[width=4.2cm]{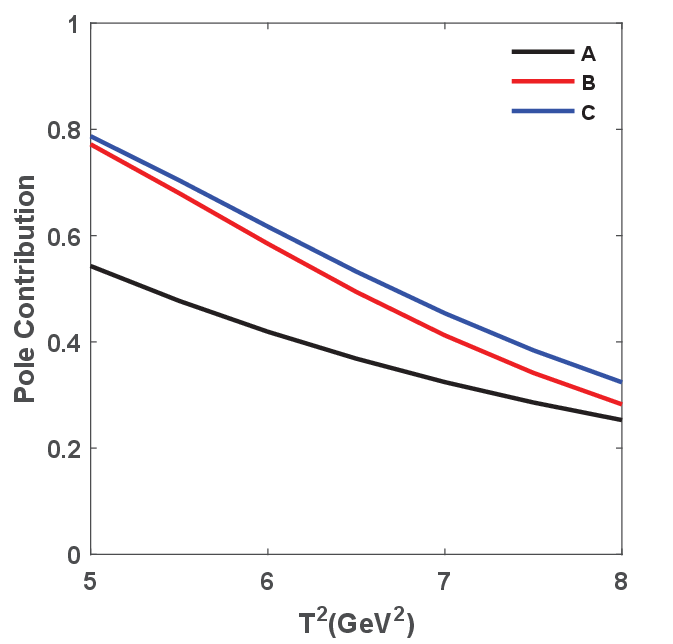}}
\subfigure[]{\includegraphics[width=4.2cm]{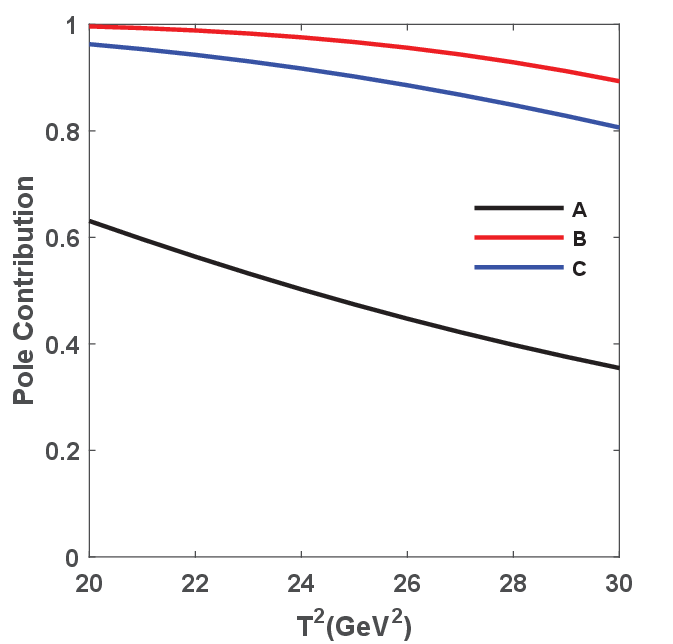}}
\caption{The pole contributions of vertices $\Sigma_{c}\Delta D^{*}$(a) and $\Sigma_{b}\Delta B^{*}$(b), where the A, B and C denote the tensor structures $i\epsilon^{\rho\tau\alpha\beta}p_{\alpha}p'_{\beta}$, $p^{\rho}p'^{\tau}$ and $p^{\rho}p^{\tau}$, respectively.}
\label{fig:PC}
\end{figure}
\begin{figure}[htbp]
\centering
\subfigure[]{\includegraphics[width=4.2cm]{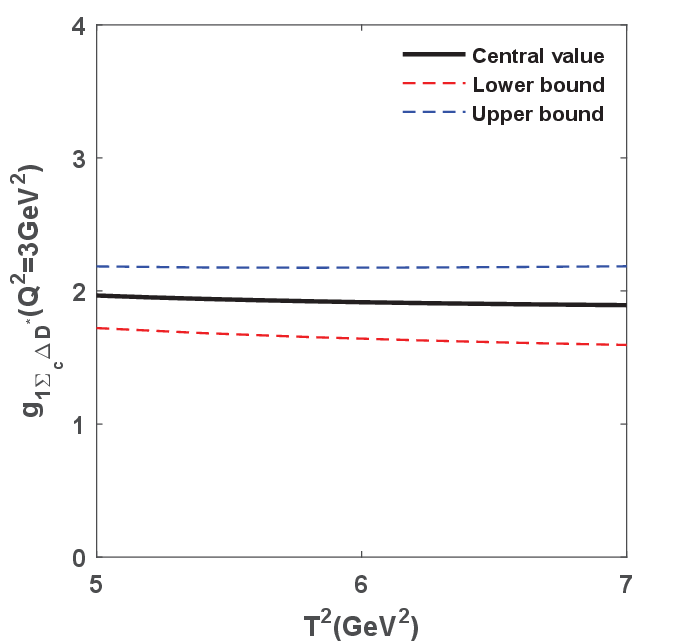}}
\subfigure[]{\includegraphics[width=4.2cm]{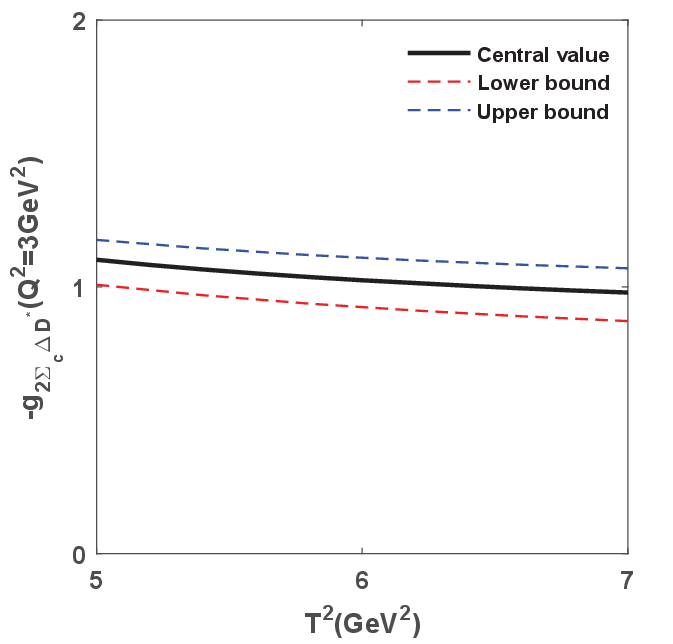}}

\subfigure[]{\includegraphics[width=4.2cm]{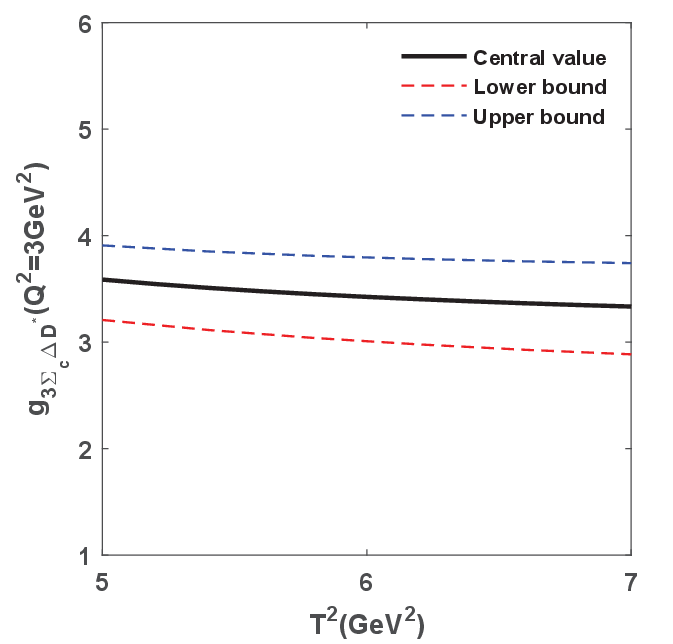}}
\subfigure[]{\includegraphics[width=4.2cm]{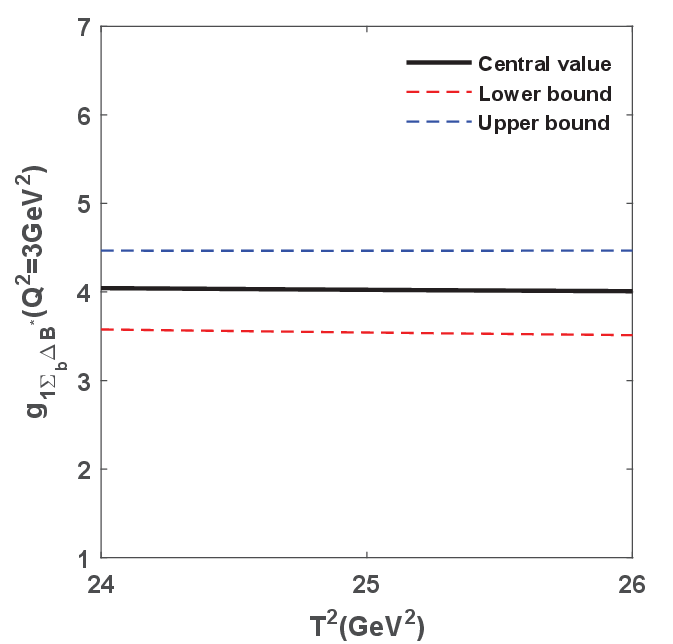}}

\subfigure[]{\includegraphics[width=4.2cm]{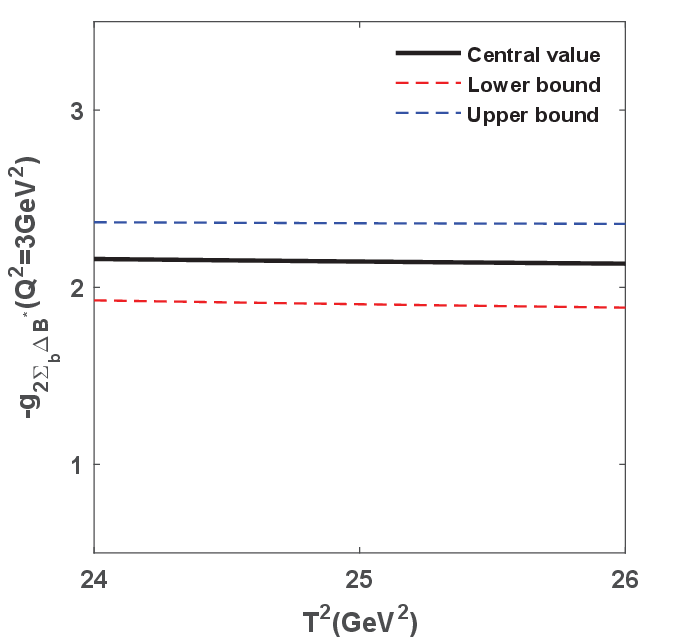}}
\subfigure[]{\includegraphics[width=4.2cm]{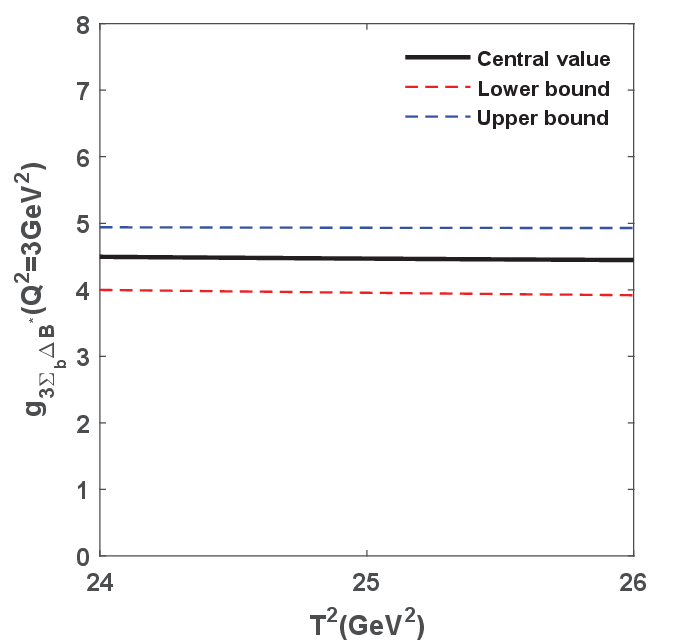}}
\caption{The coupling constants of $
g_{1\Sigma_{c}\Delta D^{*}}$(a), $-g_{2\Sigma_{c}\Delta D^{*}}$(b), $g_{3\Sigma_{c}\Delta D^{*}}$(c), and $g_{1\Sigma_{b}\Delta B^{*}}$(d), $-g_{2\Sigma_{b}\Delta B^{*}}$(e), $g_{3\Sigma_{b}\Delta B^{*}}$(f) in $Q^{2}=3$ GeV$^{2}$.}
\label{fig:BW}
\end{figure}
\begin{figure}[htbp]
\centering
\subfigure[]{\includegraphics[width=4.2cm]{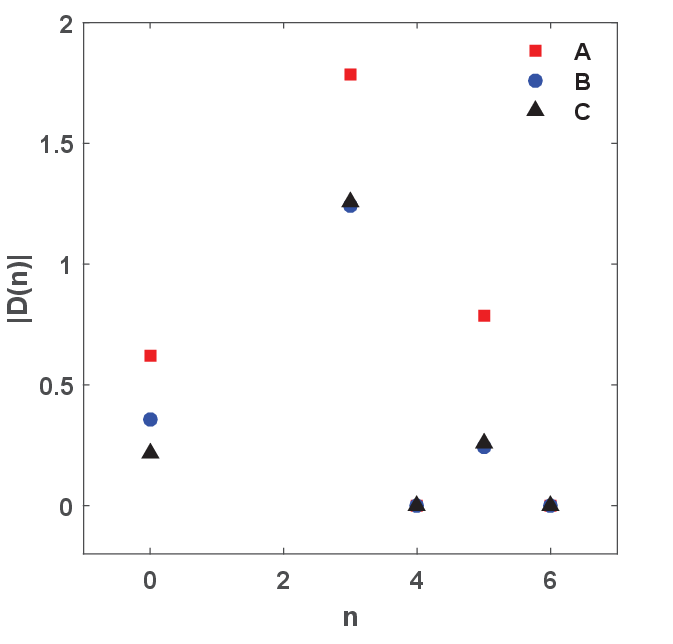}}
\subfigure[]{\includegraphics[width=4.2cm]{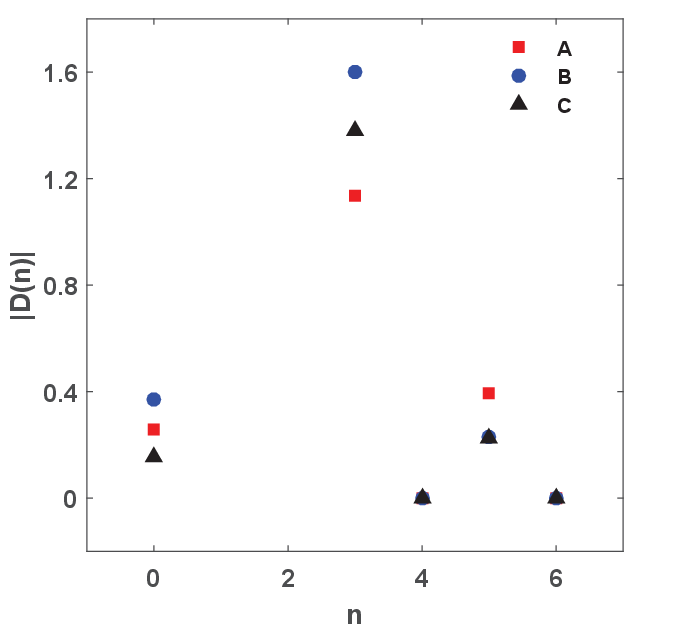}}
\caption{The dimension contributions of vertices $\Sigma_{c}\Delta D^{*}$(a) and $\Sigma_{b}\Delta B^{*}$(b), where the A, B and C denote the tensor structures $i\epsilon^{\rho\tau\alpha\beta}p_{\alpha}p'_{\beta}$, $p^{\rho}p'^{\tau}$ and $p^{\rho}p^{\tau}$, respectively.}
\label{fig:DC}
\end{figure}
\begin{figure}[htbp]
\centering
\subfigure[]{\includegraphics[width=4.2cm]{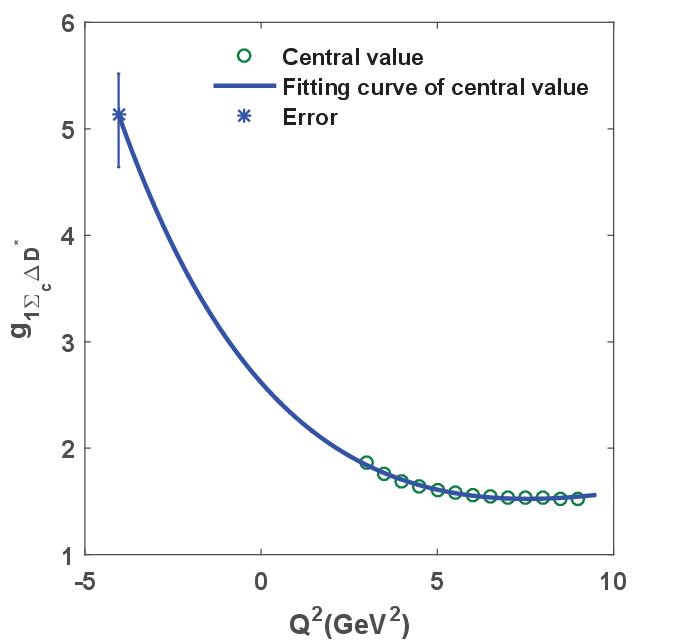}}
\subfigure[]{\includegraphics[width=4.2cm]{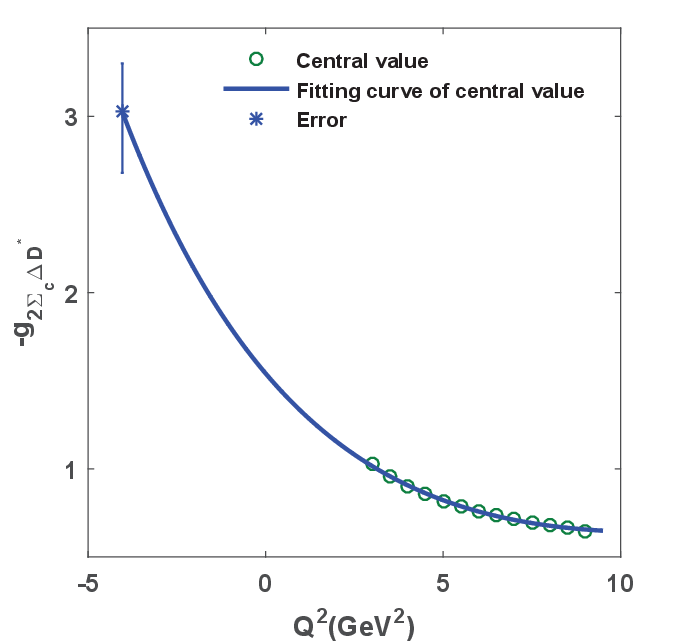}}

\subfigure[]{\includegraphics[width=4.2cm]{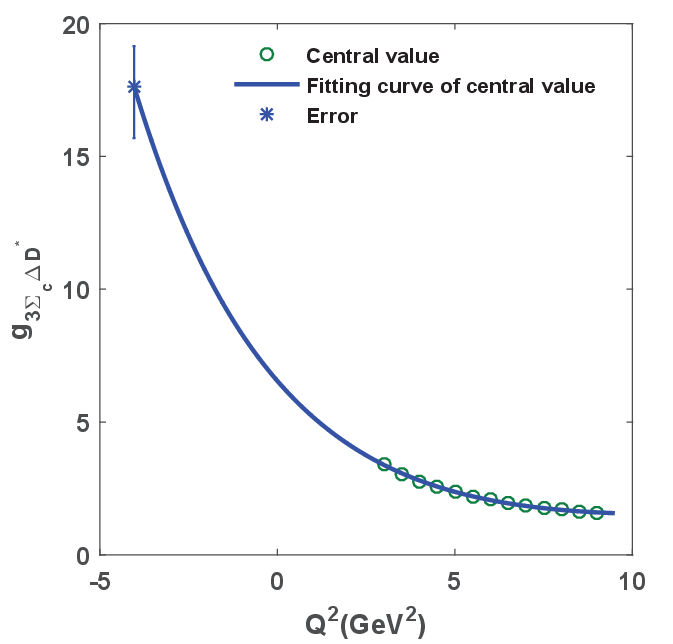}}
\subfigure[]{\includegraphics[width=4.2cm]{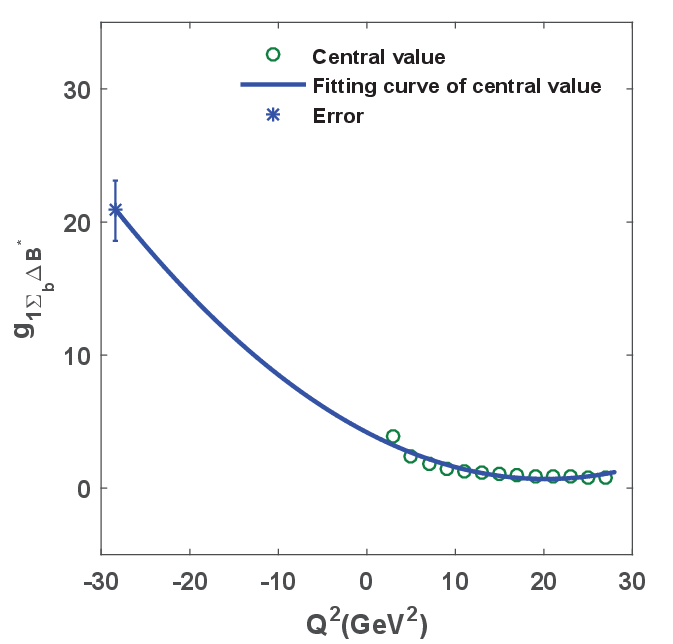}}

\subfigure[]{\includegraphics[width=4.2cm]{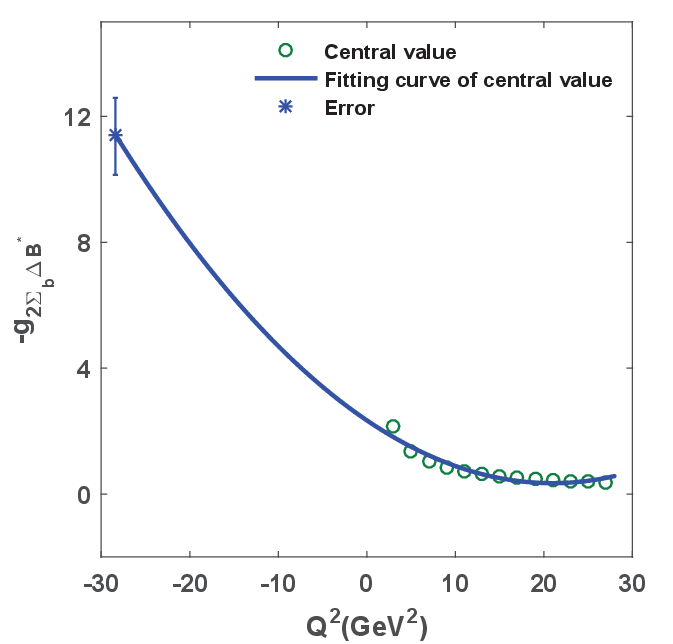}}
\subfigure[]{\includegraphics[width=4.2cm]{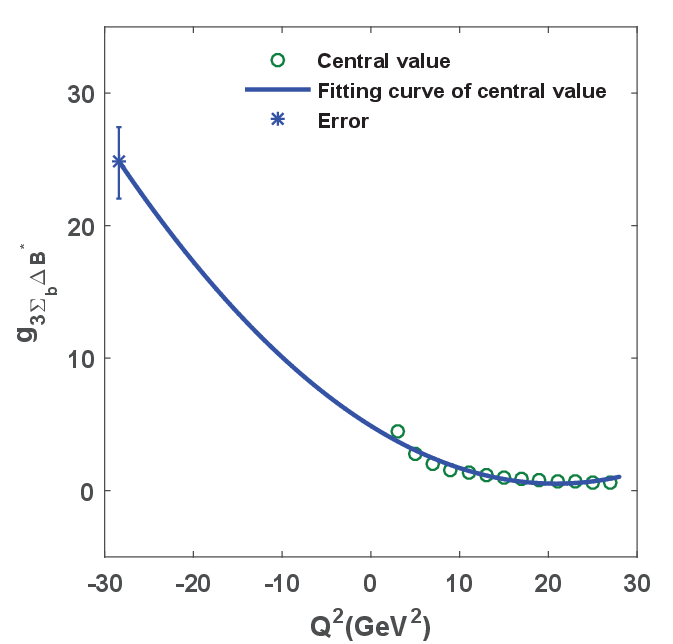}}
\caption{The fitting curves of coupling constants $
g_{1\Sigma_{c}\Delta D^{*}}$(a), $-g_{2\Sigma_{c}\Delta D^{*}}$(b), $g_{3\Sigma_{c}\Delta D^{*}}$(c), and $g_{1\Sigma_{b}\Delta B^{*}}$(d), $-g_{2\Sigma_{b}\Delta B^{*}}$(e), $g_{3\Sigma_{b}\Delta B^{*}}$(f).}
\label{fig:GG}
\end{figure}
Fixing $Q^{2}=3$ GeV$^{2}$ in Eqs. (\ref{eq:19}) and (\ref{eq:20}), we plot the pole contributions with variation of the Borel parameter for different tensor structures in Fig. \ref{fig:PC}. To satisfy the convergence of OPE, we should also find a good plateau which is generally called 'Borel window'. Then, an appropriate Borel parameter in the Borel window is selected to make pole contributions larger then 40$\%$. Considering these above requirements, the Borel windows are selected as 5(24) GeV$^2\le T^{2}\le$ 7(26) GeV$^{2}$ for the strong vertex $\Sigma_{c}\Delta D^{*}$($\Sigma_{b}\Delta B^{*}$) (see Fig. \ref{fig:BW}), Borel parameter $T^{2}$ for vertex $\Sigma_{c}\Delta D^{*}$($\Sigma_{b}\Delta B^{*}$) is taken as 6(25) GeV $^{2}$. The contributions of different vacuum condensate terms $D(n)$ are illustrated in Fig. \ref{fig:DC}, where $|D(6)|$ denotes $\langle\bar qq\rangle^{2}$ and its contribution is approximately zero. $|D(3)|$ and $|D(5)|$ which are from $\langle\bar qq\rangle$ and $\langle\overline{q}g_{s}\sigma  Gq\rangle$ satisfy $|D(3)|>|D(5)|>|D(6)|$. As for the gluon condensate $\langle g_{s}^{2}GG\rangle$, it plays a less important role since $|D(4)|<1\%$. Therefore, the convergence of OPE is well satisfied.

By taking different values of $Q^{2}$, we finally obtain the momentum dependent coupling constants $g(Q^{2})$ whose values are shown in Fig. \ref{fig:GG}. In order to obtain the on-shell values of these coupling constants, it is necessary to extrapolate these results into the time-like regions ($Q^{2}<0$). This process is realized by fitting $g(Q^{2})$ with appropriate analytical functions and setting the vector meson $D^{*}[B^{*}]$ on-shell ($Q^{2}=-m^{2}_{D^{*}[B^{*}]}$). To our knowledge, there are no specific expressions for the momentum dependent strong coupling constants which describe the interactions between hadrons. We only know that the value of running coupling constant $\alpha_{s}$(Q) decreases with the increment of square of momentum. Commonly, when we choose appropriate fitting functions, two conditions should be considered.
The first is that the coupling constants should be well fitted by the fitting functions in the space-like regions ($Q^{2}>0$). Secondly, the on-shell values of the strong coupling constants, which are obtained by extrapolating the fitting functions into deep time-like regions, should converge. Based on our previous work, the combination of exponential and polynomial functions usually satisfies these conditions. In this work, the coupling constants of vertex $\Sigma_{c}\Delta D^{*}$ are well fitted by the combination of exponential and polynomial functions. For vertex of bottom baryon, the exponential function is not well convergent in $Q^{2}=-m_{B^{*}}^{2}$ because the square mass of the vector bottom meson is much larger than that of charmed meson. Thus, the polynomial function is employed to fit the coupling constants of the vertex $\Sigma_{b}\Delta B^{*}$. Finally, the momentum dependent strong coupling constants can be fitted into the following analytical functions,

\begin{eqnarray}\label{eq22}
\notag
g_{i\Sigma_{c}\Delta D^{*}}(Q^{2})&&=aExp(-bQ^{2})+cQ^{2}  \\
g_{i\Sigma_{b}\Delta B^{*}}(Q^{2})&&=d+eQ^{2}+fQ^{4}
\end{eqnarray}
where $a$, $b$, $c$, $d$, $e$ and $f$ are the fitted parameters whose values are show in Tables~\ref{Fit1} and~\ref{Fit2}. The fitting curves for vertices $\Sigma_{c}\Delta D^{*}$ and $\Sigma_{b}\Delta B^{*}$ are also shown in Fig. \ref{fig:GG}. Finally, the on-shell values of strong coupling constants are obtained by setting $Q^{2}=-m_{D^{*}[B^{*}]}^{2}$ in Eq. (\ref{eq22}),
\begin{eqnarray}\label{eq23}
\notag
&&g_{1\Sigma_{c}\Delta D^{*}}(Q^{2}=-m_{D^{*}}^{2})=5.13^{+0.39}_{-0.49} \mathrm{GeV}^{-1}\\
\notag
&&g_{2\Sigma_{c}\Delta D^{*}}(Q^{2}=-m_{D^{*}}^{2})=-3.03^{+0.27}_{-0.35} \mathrm{GeV}^{-2}\\
\notag
&&g_{3\Sigma_{c}\Delta D^{*}}(Q^{2}=-m_{D^{*}}^{2})=17.64^{+1.51}_{-1.95} \mathrm{GeV}^{-2}\\
\notag
&&g_{1\Sigma_{b}\Delta B^{*}}(Q^{2}=-m_{B^{*}}^{2})=20.97^{+2.15}_{-2.39} \mathrm{GeV}^{-1}\\
\notag
&&g_{2\Sigma_{b}\Delta B^{*}}(Q^{2}=-m_{B^{*}}^{2})=-11.42^{+1.17}_{-1.28} \mathrm{GeV}^{-2}\\
&&g_{3\Sigma_{b}\Delta B^{*}}(Q^{2}=-m_{B^{*}}^{2})=24.87^{+2.57}_{-2.82} \mathrm{GeV}^{-2}
\end{eqnarray}
\begin{table}[htbp]
\caption{The parameters of the analytical function for the coupling constants of vertex $\Sigma_{c}\Delta D^{*}$.}
\label{Fit1}
\begin{tabular}{c c c c }
\hline\hline
Strong coupling constants&$a$&$b$&$c$ \\  \hline
$g_{1\Sigma_{c}\Delta D^{*}}$&2.618&0.189&0.119\\
$-g_{2\Sigma_{c}\Delta D^{*}}$&1.541&0.180&0.039\\
$g_{3\Sigma_{c}\Delta D^{*}}$&6.541&0.251&0.102\\
\hline\hline
\end{tabular}
\end{table}
\begin{table}[htbp]
\caption{The parameters of the analytical function for the coupling constants of vertex $\Sigma_{b}\Delta B^{*}$.}
\label{Fit2}
\begin{tabular}{c c c c}
\hline\hline
Strong coupling constants&$d$&$e$&$f$ \\  \hline
$g_{1\Sigma_{b}\Delta B^{*}}$&4.192&-0.347&0.009\\
$-g_{2\Sigma_{b}\Delta B^{*}}$&2.341&-0.191&0.005\\
$g_{3\Sigma_{b}\Delta B^{*}}$&4.894&-0.419&0.010\\
\hline\hline
\end{tabular}
\end{table}

\section{Conclusions}\label{sec4}
In this paper, we systematically analyze the strong vertices $\Sigma_{c}\Delta D^{*}$ and $\Sigma_{b}\Delta B^{*}$ using QCD sum rules, where the off-shell cases of vector mesons $D^{*}[B^{*}]$ are considered. Under this physical scheme, the momentum dependent coupling constants are obtained in the space-like ($Q^{2}>0$) regions. Then, they are fitted into analytical functions which are used to extrapolate into time-like regions($Q^{2}<0$). Finally, the on-shell values of the strong coupling constants are obtained by taking the on-shell conditions of intermediate mesons($Q^{2}=-m^{2}_{D*[B*]}$). Just as discussed in the introduction, these strong coupling constants are significant for us to understand the inner structures and strong decay behaviors of exotic hadrons.

\section*{Acknowledgements}

This project is supported by National Natural Science Foundation, Grant Number 12175068 and Natural Science Foundation of HeBei Province, Grant Number A2018502124.

\begin{widetext}
\begin{center}
\textbf{Appendix A:The calculation details of the QCD spectral density.}
\end{center}

For the perturbative part(see Fig. \ref{fig:FD}(a)), we substitute the free quark propagators in the momentum space in Eq. (\ref{eq11}). After performing integrations in the coordinate space, we can express the correlation function as follows,
\begin{eqnarray}\label{eq24}
\notag
\Pi _{\mu \nu }^{\mathrm{pert}}(p,p') &&= \frac{{12{i^2}}}{{{{\left( {2\pi } \right)}^8}}}\int {{d^4}{k_1}{d^4}{k_2}{d^4}{k_3}{d^4}{k_4}} \delta ( {p' - {k_1} - {k_2} - {k_3}})\delta ({q + {k_3} - {k_4}})\\
&&\times \Big\{ {\Big[ {\frac{{{{\slashed k}_1} + {m_q}}}{{k_1^2 - m_q^2}}}\Big]{\gamma _\alpha }\Big[{\frac{{{{\slashed k}_2} - {m_q}}}{{k_2^2 - m_q^2}}}\Big]{\gamma _\mu }\Big[{\frac{{{{\slashed k}_3} + {m_q}}}{{k_3^2 - m_q^2}}}\Big]{\gamma _\nu }\Big[{\frac{{{{\slashed k}_4} + {m_Q}}}{{k_4^2 - m_Q^2}}}\Big]{\gamma ^\alpha }{\gamma _5}}\Big\}
\end{eqnarray}
Then, we put all the quark lines on mass-shell using the Cutkosky's rules. The QCD spectral density for the perturbative part will be obtained,
\begin{eqnarray}\label{eq25}
\notag
\rho _{\mu \nu }^{\mathrm{pert}}(s,u,q^{2}) && = \frac{{12{i^2}}}{{{{\left( {2\pi } \right)}^8}}}\int {{d^4}} q'\int {{d^4}{k_3}{d^4}{k_4}} \delta ( {p' - q' - {k_3}})\delta ( {q + {k_3} - {k_4}})\\
\notag
&&\times \frac{{{{({ - 2\pi i})}^2}}}{{2\pi i}}\int_{4m_u^2}^{{{(\sqrt u  - {m_u})}^2}} {dr\frac{1}{{r - q{'^2}}}} \int {{d^4}{k_1}{d^4}{k_2}} \delta ({q' - {k_1} - {k_2}})\delta ({k_1^2 - m_q^2})\delta ({k_2^2 - m_q^2} )\\
\notag
&&\times \Big\{ {[{{{\slashed k}_1} + {m_q}}]{\gamma _\alpha }[{{{\slashed k}_2} - {m_q}}]{\gamma _\mu }\Big[ {\frac{{{{\slashed k}_3} + {m_q}}}{{k_3^2 - m_q^2}}}\Big]{\gamma _\nu }\Big[{\frac{{{{\slashed k}_4} + {m_Q}}}{{k_4^2 - m_Q^2}}}\Big]{\gamma ^\alpha }{\gamma _5}} \Big\} \\
\notag
&& \to- \frac{{12{i^2}}}{{{{\left( {2\pi } \right)}^8}}}\int {{d^4}} q'\int {{d^4}{k_3}{d^4}{k_4}} \delta( {p' - q' - {k_3}})\delta({q + {k_3} - {k_4}})\delta({k_3^2 - m_q^2})\delta({k_4^2 - m_Q^2})\delta({q{'^2}-r})\\
\notag
&&\times \frac{{{{({ - 2\pi i})}^2}}}{{2\pi i}}\frac{{{{( { - 2\pi i})}^3}}}{{(2\pi i)^{2}}}\int_{4m_u^2}^{{{(\sqrt u  - {m_u})}^2}} {dr} \int {{d^4}{k_1}{d^4}{k_2}} \delta( {q' - {k_1} - {k_2}})\delta( {k_1^2 - m_q^2})\delta ( {k_2^2 - m_q^2}) \\
\notag
&&\times \Big\{ {[{{{\slashed k}_1} + {m_q}}]{\gamma _\alpha }[ {{{\slashed k}_2} - {m_q}}]{\gamma _\mu }[{{{\slashed k}_3} + {m_q}}]{\gamma _\nu }[{{{\slashed k}_4} + {m_Q}}]{\gamma ^\alpha }{\gamma _5}} \Big\}  \\
\notag
&& = - \frac{{12{i^2}}}{{{{({2\pi })}^8}}}\int {{d^4}{k_3}}\delta({k_3^2 - m_q^2})\delta([{k_3+p-p'}]^2-m_Q^2)\delta([{k_3-p']^2 - r})\\
\notag
&&\times \frac{{{{({ - 2\pi i})}^2}}}{{2\pi i}}\frac{{{{( { - 2\pi i})}^3}}}{{(2\pi i)^{2}}}\int_{4m_u^2}^{{{(\sqrt u  - {m_u})}^2}} {dr} \int {{d^4}{k_1}}\delta( {k_1^2 - m_q^2})\delta ([{q'-k_1}]^2 - m_q^2) \\
\notag
&&\times \Big\{ {[{{{\slashed k}_1} + {m_q}}]{\gamma _\alpha }[{{{\slashed k}_2} - {m_q}}]{\gamma _\mu }[{{{\slashed k}_3} + {m_q}}]{\gamma _\nu }[{{{\slashed k}_4} + {m_Q}}]{\gamma ^\alpha }{\gamma _5}}\Big\} \\
\notag
&& =\frac{12}{{{{({2\pi })}^6}}}\frac{\pi}{2\sqrt{\lambda(s,u,q^{2})}}\int_{4m_u^2}^{{{(\sqrt u  - {m_u})}^2}} {dr}\frac{\pi\sqrt{\lambda(r,m_{q}^{2},m_{q}^{2})}}{2r}\Big\{[{\frac{1}{2}(\slashed p'-C_{p'}\slashed p'-C_p\slashed p) + {m_q}}]{\gamma _\alpha } \\
&&\times [{\frac{1}{2}(\slashed p'-C_{p'}\slashed p'-C_p\slashed p)  {m_q}}]{\gamma _\mu }[{C_{p}\slashed p+C_{p'}\slashed p' + {m_q}}]{\gamma _\nu }[{(C_{p}+1)\slashed p+(C_{p'}-1)\slashed p' + {m_Q}}]{\gamma ^\alpha }{\gamma _5} \Big\}
\end{eqnarray}
where,
\begin{eqnarray}
\notag
&&C_{p}=\frac{(u+m_{q}^{2}-r)(s+u-q^{2})-2u(u-q^2+m_{Q}^{2}-r)}{\lambda(s,u,q^{2})} \\
\notag
&&C_{p'}=\frac{(u-q^{2}+m_{Q}^{2}-r)(s+u-q^{2})-2s(u+m_{q}^{2}-r)}{\lambda(s,u,q^{2})} \\
&&\lambda(a,b,c)=a^{2}+b^{2}+c^{2}-2ab-2bc-2ac
\end{eqnarray}

The non-perturbative terms including $\langle\overline{q}q\rangle$, $\langle g_{s}^{2}GG \rangle$, $\langle\overline{q}g_{s}\sigma  Gq\rangle$ and $\langle\overline{q}q\rangle^{2}$ are also calculated by using the Cutkosky's rules. For the condensate terms $\langle\overline{q}q\rangle$, $\langle\overline{q}g_{s}\sigma  Gq\rangle$ and $\langle\overline{q}q\rangle^{2}$, their correlation functions can be expressed as,
\begin{eqnarray}
\notag
{\Pi _{\mu \nu }}(p,p') &&= \frac{{12i}}{{{{({2\pi })}^4}}}\int {{d^4}k\frac{1}{{[ {{{( {k - p'})}^2} - m_q^2}]({{k^2} - m_q^2})[{{{({p - p' + k})}^2} - m_Q^2}]}}} \\
\notag
&& \times \Big\{ [ - \frac{{\left\langle {\overline q q} \right\rangle }}{{12}} + \frac{{\left\langle {\overline q {g_s}\sigma Gq} \right\rangle }}{{192}}\partial _{p'}^2 + \frac{{{m_q}\left\langle {\overline q q} \right\rangle }}{{48}}{{\slashed \partial }_{p'}} - \frac{{{m_q}\left\langle {\overline q {g_s}\sigma Gq} \right\rangle }}{{1152}}\partial _{p'}^2{{\slashed \partial }_{p'}}\\
\notag
&& + \frac{{g_s^2{{\left\langle {\overline q q} \right\rangle }^2}}}{{7776}}\partial _{p'}^2{{\slashed \partial }_{p'}}]{\gamma _\alpha }\left( {\slashed p' - \slashed k - {m_q}} \right){\gamma _\mu }\left( {\slashed k + {m_q}} \right){\gamma _\nu }\left( {\slashed k + \slashed q + {m_Q}} \right){\gamma ^\alpha }{\gamma _5}\\
\notag
&& + (\slashed p' - \slashed k + {m_q}){\gamma _\alpha }[\frac{{\left\langle {\overline q q} \right\rangle }}{{12}} - \frac{{\left\langle {\overline q {g_s}\sigma Gq} \right\rangle }}{{192}}\partial _{p'}^2 + \frac{{{m_q}\left\langle {\overline q q} \right\rangle }}{{48}}{{\slashed \partial }_{p'}} - \frac{{{m_q}\left\langle {\overline q {g_s}\sigma Gq} \right\rangle }}{{1152}}\partial _{p'}^2{{\slashed \partial }_{p'}}\\
&& + \frac{{g_s^2{{\left\langle {\overline q q} \right\rangle }^2}}}{{7776}}\partial _{p'}^2{{\slashed \partial }_{p'}}]{\gamma _\mu }(\slashed k + {m_q}){\gamma _\nu }(\slashed q + {{\slashed k}} + {m_Q}){\gamma ^\alpha }{\gamma _5}\Big\}
\end{eqnarray}
The QCD spectral density can also be obtained by the Cutkosky's rules,
\begin{eqnarray}
\notag
{\rho _{\mu \nu }}(p,p') &&= \frac{{12i}}{{{{\left( {2\pi } \right)}^4}}}\frac{{{{( - 2\pi i)}^3}}}{{{{(2\pi i)}^2}}}\int {{d^4}k\delta [{{\left( {k - p'} \right)}^2} - m_q^2]\delta [{k^2} - m_q^2]\delta [{{\left( {p - p' + k} \right)}^2} - m_Q^2} ]\\
\notag
&& \times \Big\{ \Big[ - \frac{{\left\langle {\overline q q} \right\rangle }}{{12}} + \frac{{\left\langle {\overline q {g_s}\sigma Gq} \right\rangle }}{{192}}\partial _{p'}^2 + \frac{{{m_q}\left\langle {\overline q q} \right\rangle }}{{48}}{{\slashed \partial }_{p'}} - \frac{{{m_q}\left\langle {\overline q {g_s}\sigma Gq} \right\rangle }}{{1152}}\partial _{p'}^2{{\slashed \partial }_{p'}}\\
\notag
&& + \frac{{g_s^2{{\left\langle {\overline q q} \right\rangle }^2}}}{{7776}}\partial _{p'}^2{{\slashed \partial }_{p'}}\Big]{\gamma _\alpha }(\slashed p' - \slashed k + {m_q}){\gamma _\mu }(\slashed k + {m_q}){\gamma _\nu }(\slashed q + {{\slashed k}_3} + {m_Q}){\gamma ^\alpha }{\gamma _5}\\
\notag
&& + (\slashed p' - \slashed k + {m_q}){\gamma _\alpha }\Big[\frac{{\left\langle {\overline q q} \right\rangle }}{{12}} - \frac{{\left\langle {\overline q {g_s}\sigma Gq} \right\rangle }}{{192}}\partial _{p'}^2 + \frac{{{m_q}\left\langle {\overline q q} \right\rangle }}{{48}}{{\slashed \partial }_{p'}} - \frac{{{m_q}\left\langle {\overline q {g_s}\sigma Gq} \right\rangle }}{{1152}}\partial _{p'}^2{{\slashed \partial }_{p'}}\\
\notag
&& + \frac{{g_s^2{{\left\langle {\overline q q} \right\rangle }^2}}}{{7776}}\partial _{p'}^2{{\slashed \partial }_{p'}}\Big]{\gamma _\mu }(\slashed k + {m_q}){\gamma _\nu }(\slashed q + {{\slashed k}_3} + {m_Q}){\gamma ^\alpha }{\gamma _5}\Big\} \\
\notag
&& = \frac{{12}}{{{{(2\pi )}^3}}}\frac{\pi }{{2\sqrt {\lambda (s,u,{q^2})} }}\Big\{ \Big[ - \frac{{\left\langle {\overline q q} \right\rangle }}{{12}} + \frac{{\left\langle {\overline q {g_s}\sigma Gq} \right\rangle }}{{192}}\partial _{p'}^2 + \frac{{{m_q}\left\langle {\overline q q} \right\rangle }}{{48}}{{\slashed \partial }_{p'}} - \frac{{{m_q}\left\langle {\overline q {g_s}\sigma Gq} \right\rangle }}{{1152}}\partial _{p'}^2{{\slashed \partial }_{p'}}\\
\notag
&& + \frac{{g_s^2{{\left\langle {\overline q q} \right\rangle }^2}}}{{7776}}\partial _{p'}^2{{\slashed \partial }_{p'}}\Big]{\gamma _\alpha }[(1 - C{'_{p'}})\slashed p' - C{'_p}\slashed{p} - {m_q}]{\gamma _\mu }[C{'_p}\slashed{p} + C{'_{p'}}\slashed{p}' + {m_q}]{\gamma _\nu }[(C{'_p} + 1)\slashed{p} + (C{'_{p'}} - 1)\slashed{p}' + {m_Q}]{\gamma ^\alpha }{\gamma _5}\\
\notag
&& + [(1 - C{'_{p'}})\slashed p' - C{'_p}\slashed{p} + {m_q}]{\gamma _\alpha }\Big[\frac{{\left\langle {\overline q q} \right\rangle }}{{12}} - \frac{{\left\langle {\overline q {g_s}\sigma Gq} \right\rangle }}{{192}}\partial _{p'}^2 + \frac{{{m_q}\left\langle {\overline q q} \right\rangle }}{{48}}{{\slashed \partial }_{p'}} - \frac{{{m_q}\left\langle {\overline q {g_s}\sigma Gq} \right\rangle }}{{1152}}\partial _{p'}^2{{\slashed \partial }_{p'}}\\
&& + \frac{{g_s^2{{\left\langle {\overline q q} \right\rangle }^2}}}{{7776}}\partial _{p'}^2{{\slashed \partial }_{p'}}\Big]{\gamma _\mu }[C{'_p}\slashed{p} + C{'_{p'}}\slashed{p}' + {m_q}]{\gamma _\nu }[(C{'_p} + 1)\slashed{p} + (C{'_{p'}} - 1)\slashed{p}' + {m_Q}]{\gamma ^\alpha }{\gamma _5}\Big\}
\end{eqnarray}
where,
\begin{eqnarray}
\notag
C{'_p} &&= \frac{{u\left( {s + u - {q^2}} \right) - 2u(u - {q^2} + m_Q^2 - m_q^2)}}{{\lambda \left( {s,u,{q^2}} \right)}}\\
C{'_{p'}} &&= \frac{{\left( {u - {q^2} + m_Q^2 - m_q^2} \right)\left( {s + u - {q^2}} \right) - 2su}}{{\lambda \left( {s,u,{q^2}} \right)}}
\end{eqnarray}

As for the gluon condensate, a typical integral will be encountered,
\begin{eqnarray}
{I_{ijkl}} = \int {{d^4}{k_1}} {d^4}{k_2}{d^4}{k_3}{d^4}{k_4}\frac{1}{{{{(k_1^2 - m_1^2)}^i}{{(k_2^2 - m_2^2)}^j}{{(k_3^2 - m_3^2)}^k}{{(k_4^2 - m_4^2)}^l}}}
\end{eqnarray}
According to the following transformation, this terms can also be calculated,
\begin{eqnarray}
\notag
{I_{ijkl}} &&= \frac{1}{{(i - 1)!(j - 1)!(k - 1)!(l - 1)!}}\frac{\partial^{i-1}}{{\partial A^{i-1}}}\frac{\partial^{j-1}}{{\partial B^{j-1}}}\frac{\partial^{k-1}}{{\partial C^{k-1}}}\frac{\partial^{l-1}}{{\partial D^{l-1}}}\int {{d^4}{k_1}} {d^4}{k_2}{d^4}{k_3}{d^4}{k_4}\\
\notag
&&\times \frac{1}{{(k_1^2 - A)(k_2^2 - B)(k_3^2 - C)(k_4^2 - D)}}{|_{_{A \to m_1^2,B \to m_2^2,C \to m_3^2,D \to m_4^2}}}\\
\notag
&& \to \frac{{{{( - 2\pi i)}^4}}}{{(i - 1)!(j - 1)!(k - 1)!(l - 1)!}}\frac{\partial^{i-1}}{{\partial A^{i-1}}}\frac{\partial^{j-1}}{{\partial B^{j-1}}}\frac{\partial^{k-1}}{{\partial C^{k-1}}}\frac{\partial^{l-1}}{{\partial D^{l-1}}} \int {{d^4}{k_1}} {d^4}{k_2}{d^4}{k_3}{d^4}{k_4}\delta (k_1^2 - A)\delta (k_2^2 - B)\delta (k_3^2 - C)\\
&&\times\delta (k_4^2 - D){|_{_{A \to m_1^2,B \to m_2^2,C \to m_3^2,D \to m_4^2}}}
\end{eqnarray}
\end{widetext}

\begin{widetext}
\begin{center}
\textbf{Appendix B:Full expressions of the QCD spectral density.}
\end{center}
\begin{eqnarray}
\notag
\bar \rho _1^{pert}(s,u,r,{Q^2}) &&= \frac{3}{{32{\pi ^4}{{[{Q^4} + 2{Q^2}(s + u) + {{(s - u)}^2}]}^{\frac{5}{2}}}}}\{ m_Q^6u({Q^2} + s - u) + m_Q^4\{ {Q^4}(r + 2u) + {Q^2}[2r(s - 2u) + u(s + u)]\\
\notag
&& + (r - u){(s - u)^2}\}  + m_Q^2\{ {Q^6}(r + u) + {Q^4}[ - 3{r^2} + r(s - 3u) + u(2u - s)] - {Q^2}(s - u)[3{r^2} + r(s - 3u)\\
\notag
&& + u(2s + u)] - r{(s - u)^3}\}  - {Q^6}(r + s)(r + u) - {Q^4}[ - 2{r^3} - {r^2}(s + u) + 2r({s^2} - su + {u^2}) + su(s + u)]\\
&& + {Q^2}r{(s - u)^2}(2r - s - u)\}
\end{eqnarray}
\begin{eqnarray}
\notag
\bar \rho _2^{pert}(s,u,r,{Q^2}) &&=  - \frac{3}{{32{\pi ^4}{Q^2}{{[{Q^4} + 2{Q^2}(s + u) + {{(s - u)}^2}]}^{\frac{7}{2}}}}}\{  - m_Q^4u - m_Q^2[{Q^2}(r + u) + (r - u)(s - u)] + {Q^2}(r + s)(r + u)\\
\notag
&& + r{(s - u)^2}\} \{ m_Q^4u[ - 3{Q^4} - 2{Q^2}(s + u) + {(s - u)^2}] + m_Q^2{Q^2}[{Q^4} + 2{Q^2}(s - u) + {s^2} + 2su\\
\notag
&& - 3{u^2}]({Q^2} - 2r + s + u) + {Q^2}\{ {Q^8} + 3{Q^6}( - r + s + u) + {Q^4}(3{r^2} - 5sr - 5ur + 3{s^2} + 5su + 3{u^2})\\
&& + {Q^2}[2{r^2}(s + u) - r({s^2} + 6su + {u^2}) + {s^3} + {s^2}u + s{u^2} + {u^3}] - (r - s)(r - u){(s - u)^2}\} \}
\end{eqnarray}
\begin{eqnarray}
\notag
\bar \rho _3^{pert}(s,u,r,{Q^2})&& = \frac{3}{{32{\pi ^4}{Q^2}{{[{Q^4} + 2{Q^2}(s + u) + {{(s - u)}^2}]}^{\frac{7}{2}}}}}\{  - m_Q^4u - m_Q^2[{Q^2}(r + u) + (r - u)(s - u)] + {Q^2}(r + s)(r + u)\\
\notag
&& + r{(s - u)^2})\} \{ m_Q^4u{(s - u)^2} + {Q^6}[u(4m_Q^2 + 3s + 4u) - {r^2} + r(3s - 5u)] + {Q^4}[r( - 8um_Q^2 - 5{u^2}\\
\notag
&& + 3{s^2} - 6su) + u(m_Q^4 + 8m_Q^2s + 2us + 2{u^2}) - 2{r^2}(s - 3u)] - {Q^2}[ - r(s - u)({s^2} - 8um_Q^2 - {u^2})\\
&& + u( - 2sm_Q^4 + 6um_Q^4 - 4{s^2}m_Q^2 + 4{u^2}m_Q^2 + {s^3} - 2{s^2}u + s{u^2}) + {r^2}{(s - u)^2}] + {Q^8}(r + 2u)\}
\end{eqnarray}
\begin{eqnarray}
\bar \rho _1^{\left\langle {\bar qq} \right\rangle }(s,u,{Q^2}) &&= \frac{{3{m_Q}\left\langle {\bar qq} \right\rangle u}}{{2{\pi ^2}{{[{Q^4} + 2{Q^2}(s + u) + {{(s - u)}^2}]}^{\frac{3}{2}}}}}(2m_Q^2 + {Q^2} - s + u) \\ \notag \\
\notag
\bar \rho _2^{\left\langle {\bar qq} \right\rangle }(s,u,{Q^2})&& =  - \frac{{3{m_Q}\left\langle {\bar qq} \right\rangle u}}{{2{\pi ^2}{Q^2}{{[{Q^4} + 2{Q^2}(s + u) + {{(s - u)}^2}]}^{\frac{5}{2}}}}}\{ m_Q^4[3{Q^4} + 2{Q^2}(s + u) - {(s - u)^2}] + m_Q^2[3{Q^6} - {Q^4}(s - 5u)\\
&& + {Q^2}( - 3{s^2} + 2su + {u^2}) + {(s - u)^3}] + {Q^2}[{Q^6} + 3{Q^4}u + {Q^2}({s^2} + 3{u^2}) + {(s - u)^2}(2s + u)]\}
\end{eqnarray}
\begin{eqnarray}
\notag
\bar \rho _3^{\left\langle {\bar qq} \right\rangle }(s,u,{Q^2}) &&= \frac{{3{m_Q}\left\langle {\bar qq} \right\rangle u}}{{2{\pi ^2}{Q^2}{{[{Q^4} + 2{Q^2}(s + u) + {{(s - u)}^2}]}^{\frac{5}{2}}}}}\{ m_Q^4[ - {Q^4} - 2{Q^2}(s - 3u) - {(s - u)^2}] - m_Q^2[{Q^6}+ {Q^4}(s - 5u)\\
&& - {Q^2}({s^2} - 6su + 5{u^2}) - {(s - u)^3}] + {Q^2}[{Q^4}(s + 2u) + 2{Q^2}({s^2} - su + 2{u^2}) + {(s - u)^2}(s + 2u)]\}
\end{eqnarray}
\begin{eqnarray}
\bar \rho _1^{\left\langle {\bar q{g_s}\sigma Gq} \right\rangle }(s,u,{Q^2})&& =  - \frac{{3{m_Q}\left\langle {\bar q{g_s}\sigma Gq} \right\rangle }}{{8{\pi ^2}{{[{Q^4} + 2{Q^2}(s + u) + {{(s - u)}^2}]}^{\frac{3}{2}}}}}({Q^2} + s + u)\\ \notag \\
\notag
\bar \rho _2^{\left\langle {\bar q{g_s}\sigma Gq} \right\rangle }(s,u,{Q^2})&& = \frac{{3{m_Q}\left\langle {\bar q{g_s}\sigma Gq} \right\rangle }}{{8{\pi ^2}{{[{Q^4} + 2{Q^2}(s + u) + {{(s - u)}^2}]}^{\frac{5}{2}}}}}\{ m_Q^4({Q^4} + 2{Q^2}u - {s^2} - 2su + {u^2}) + m_Q^2[3{Q^6} + {Q^4}(3s + 5u)\\
&& + {Q^2}({s^2} + 4su + {u^2}) + {s^3} - {s^2}u + s{u^2} - {u^3}] + ({Q^2} + u)[{Q^6} + {Q^4}u - {Q^2}({s^2} - 4su + {u^2}) - u{(s - u)^2}]\}
\end{eqnarray}
\begin{eqnarray}
\notag
\bar \rho _3^{\left\langle {\bar q{g_s}\sigma Gq} \right\rangle }(s,u,{Q^2}) &&= \frac{{3{m_Q}\left\langle {\bar q{g_s}\sigma Gq} \right\rangle }}{{8{\pi ^2}{{[{Q^4} + 2{Q^2}(s + u) + {{(s - u)}^2}]}^{\frac{5}{2}}}}}\{ (m_Q^2 - s - 3u){({Q^2} + s + u)^3} + [m_Q^4 - 2m_Q^2(s + 4u) + {s^2}+ 14su\\
\notag
&& + 4{u^2}]{({Q^2} + s + u)^2} - 2u[m_Q^4 - m_Q^2(7s + 4u) + 3s(2s + 3u)]({Q^2} + s + u)\\
&& + 2su[m_Q^4 - 2m_Q^2(s + 2u) + {s^2} + 5su + 2{u^2}]\}
\end{eqnarray}
\begin{eqnarray}
\notag
\bar \rho _1^{{{\left\langle {\bar qq} \right\rangle }^2}}(s,u,{Q^2})&& =  - \frac{{sg_s^2{{\left\langle {\bar qq} \right\rangle }^2}}}{{81{\pi ^2}{{[{Q^4} + 2{Q^2}(s + u) + {{(s - u)}^2}]}^{\frac{7}{2}}}}}\{ 2m_Q^4[{Q^4} + 2{Q^2}(s + u) + {s^2} + 3su + {u^2}] + m_Q^2[{Q^6}-{Q^4}(s - 3u) \\
\notag
&& + {Q^2}( - 5{s^2} + 4su + 3{u^2}) - 3{s^3} - 3{s^2}u + 5s{u^2} + {u^3}] + s[ - {Q^6} - {Q^4}(s + u) + {Q^2}({s^2} - 4su + {u^2})\\
&& + {(s - u)^2}(s + u)]\}
\end{eqnarray}
\begin{eqnarray}
\notag
\bar \rho _2^{{{\left\langle {\bar qq} \right\rangle }^2}}(s,u,{Q^2})&& = \frac{{g_s^2{{\left\langle {\bar qq} \right\rangle }^2}}}{{162{\pi ^2}{Q^2}{{[{Q^4} + 2{Q^2}(s + u) + {{(s - u)}^2}]}^{\frac{9}{2}}}}}{\rm{\{ }}6{Q^{16}} + 4(9s + 10u){Q^{14}} + (63{s^2} + 152us + 112{u^2}){Q^{12}}\\
\notag
&& + 2(7{s^3} + 101u{s^2} + 116{u^2}s + 84{u^3}){Q^{10}} - 7(9{s^4} - 16u{s^3} - 33{u^2}{s^2} - 20{u^3}s - 20{u^4}){Q^8} + ( - 56{s^5} + 24u{s^4}\\
\notag
&& + 40{u^2}{s^3} + 84{u^3}{s^2} + 20{u^4}s + 56{u^5}){Q^6} + s( - 7{s^5} - 8u{s^4} + 18{u^2}{s^3} + 40{u^3}{s^2} - 59{u^4}s + 16{u^5}){Q^4}\\
\notag
&& + 2{(s - u)^3}(3{s^4} + 15{u^2}{s^2} - 4{u^3}s + 4{u^4}){Q^2} + {(s - u)^5}({s^3} - 3u{s^2} - 2{u^2}s + 2{u^3}) - m_Q^6[9{Q^{10}}\\
\notag
&& + (37s + 9u){Q^8} + (58{s^2} + 94us - 54{u^2}){Q^6} + 6(7{s^3} + 30u{s^2} + 20{u^2}s - 21{u^3}){Q^4} + (13{s^4}\\
\notag
&& + 114u{s^3} + 138{u^2}{s^2} + 106{u^3}s - 99{u^4}){Q^2} + {(s - u)^2}({s^3} + 21u{s^2} - 11{u^2}s - 27{u^3})] + m_Q^2[ - {Q^{14}}\\
\notag
&& + (41s - 5u){Q^{12}} + 3(52{s^2} + 26us - 3{u^2}){Q^{10}} + (194{s^3} + 282u{s^2} - 73{u^2}s - 5{u^3}){Q^8} + (71{s^4}\\
\notag
&& + 298u{s^3} + 182{u^2}{s^2} - 252{u^3}s + 5{u^4}){Q^6} + ( - 27{s^5} + 81u{s^4} + 80{u^2}{s^3} + 26{u^3}{s^2} - 153{u^4}s + 9{u^5}){Q^4}\\
\notag
&& - {(s - u)^2}(18{s^4} + 60u{s^3} + 103{u^2}{s^2} - 24{u^3}s - 5{u^4}){Q^2} - {(s - u)^4}u(6{s^2} - 29us - {u^2})] + m_Q^4[ - 19{Q^{12}}\\
\notag
&& - 2(25s + 33u){Q^{10}} - (9{s^2} + 212us + 45{u^2}){Q^8} + 4(21{s^3} - 34u{s^2} - 96{u^2}s + 25{u^3}){Q^6} + (91{s^4} + 120u{s^3}\\
\notag
&& - 178{u^2}{s^2} - 380{u^3}s + 195{u^4}){Q^4} + 2(15{s^5} + 65u{s^4} + 12{u^2}{s^3} - 52{u^3}{s^2} - 103{u^4}s + 63{u^5}){Q^2}\\
&& + {(s - u)^3}({s^3} + 23u{s^2} - 39{u^2}s - 29{u^3})]{\rm{\} }}
\end{eqnarray}
\begin{eqnarray}
\notag
\bar \rho _3^{{{\left\langle {\bar qq} \right\rangle }^2}}(s,u,{Q^2}) &&= \frac{{g_s^2{{\left\langle {\bar qq} \right\rangle }^2}}}{{162{\pi ^2}{Q^2}{{[{Q^4} + 2{Q^2}(s + u) + {{(s - u)}^2}]}^{\frac{9}{2}}}}}\{ m_Q^6[{Q^{10}} + {Q^8}(5s - 23u) + 2{Q^6}(5{s^2} - 25su- 31{u^2}) \\
\notag
&& + 2{Q^4}(5{s^3} - 6{s^2}u - 108s{u^2} - 19{u^3}) + {Q^2}(5{s^4} + 34{s^3}u - 150{s^2}{u^2} - 142s{u^3} + 13{u^4})\\
\notag
&& + {(s - u)^2}({s^3} + 21{s^2}u + 45s{u^2} + 13{u^3})] + m_Q^4[{Q^{12}} + 4{Q^{10}}(s + u) + {Q^8}(5{s^2} + 116su + 5{u^2}) + 300{Q^6}{s^2}u\\
\notag
&& + {Q^4}( - 5{s^4} + 244{s^3}u + 342{s^2}{u^2} - 296s{u^3} - 5{u^4}) - 4{Q^2}({s^5} - 8{s^4}u - 98{s^3}{u^2} + 67{s^2}{u^3} + 37s{u^4} + {u^5})\\
\notag
&& - {(s - u)^3}({s^3} + 27{s^2}u + 33s{u^2} - {u^3})] - m_Q^2[{Q^{14}} + {Q^{12}}(9s - 3u) + {Q^{10}}(32{s^2} - 16su - 27{u^2})\\
\notag
&& + {Q^8}(60{s^3} + 80{s^2}u - 67s{u^2} - 55{u^3}) + {Q^6}(65{s^4} + 298{s^3}u - 366{s^2}{u^2} - 24s{u^3} - 45{u^4}) + {Q^4}(41{s^5}\\
\notag
&& + 299{s^4}u - 288{s^3}{u^2} - 190{s^2}{u^3} + 35s{u^4} - 9{u^5}) + {Q^2}{(s - u)^2}(14{s^4} + 114{s^3}u + 267{s^2}{u^2} + 22s{u^3} + 7{u^4})\\
\notag
&& + {(s - u)^4}(2{s^3} + 3s{u^2} + 3{u^3})] + s[{Q^{12}}(s - 4u) + 2{Q^{10}}s(3s - 13u) + {Q^8}(15{s^3} - 28{s^2}u + 21s{u^2} + 36{u^3})\\
\notag
&& + 4{Q^6}(5{s^4} + 7{s^3}u - 36{s^2}{u^2} + 14s{u^3} + 16{u^4}) + {Q^4}(15{s^5} + 52{s^4}u - 222{s^3}{u^2} + 160{s^2}{u^3} - 41s{u^4} + 36{u^5})\\
&& + 2{Q^2}s{(s - u)^3}(3{s^2} + 16su + 15{u^2}) + {(s - u)^5}({s^2} + su + 4{u^2})]\}
\end{eqnarray}
\begin{eqnarray}
\bar \rho _1^{\left\langle {g_s^2GG} \right\rangle }(s,u,r,{Q^2})&& = \frac{{m_Q^2\left\langle {g_s^2GG} \right\rangle u}}{{128{\pi ^4}{{[{Q^4} + 2{Q^2}(s + u) + {{(s - u)}^2}]}^{\frac{5}{2}}}}}({Q^2} + s - u)\\
\notag
\bar \rho _2^{\left\langle {g_s^2GG} \right\rangle }(s,u,r,{Q^2})&& = \frac{{m_Q^2\left\langle {g_s^2GG} \right\rangle u}}{{128{\pi ^4}{Q^2}{{[{Q^4} + 2{Q^2}(s + u) + {{(s - u)}^2}]}^{\frac{7}{2}}}}}\{ {Q^4}[ - 2u(6m_Q^2 + 5u) + r(5u - 9s) + 3{s^2} + 3su]+ {Q^2}[ - su\\
\notag
&&\times (8m_Q^2 + 3u) - 4{u^2}(2m_Q^2 + u) - 3r({s^2} + 2su - 3{u^2}) + {s^3} + 6{s^2}u] + {(s - u)^2}[u(4m_Q^2 - s+ u) + r(s - u)]\\
&& + {Q^8} + {Q^6}( - 5r + 3s - 4u)\}
\end{eqnarray}
\begin{eqnarray}
\notag
\bar \rho _3^{\left\langle {g_s^2GG} \right\rangle }(s,u,r,{Q^2}) &&=  - \frac{{m_Q^2\left\langle {g_s^2GG} \right\rangle u}}{{128{\pi ^4}{Q^2}{{[{Q^4} + 2{Q^2}(s + u) + {{(s - u)}^2}]}^{\frac{7}{2}}}}}\{ {Q^4}[u(4m_Q^2 + 9s - 5u) + 3r(s - 5u)]+ {Q^2}[8m_Q^2u(s - 3u)\\
&& + 3u({s^2} + 2su - 3{u^2}) + 3r({s^2} - 6su + 5{u^2})] + {(s - u)^2}[u(4m_Q^2 - s + u)+ r(s - u)] + {Q^6}(r + 5u)\}
\end{eqnarray}
\end{widetext}

\end{document}